\documentclass[aps,twocolumn,showpacs,email,amssymb,nobibnotes]{revtex4-1}

\usepackage{bm}
\usepackage{graphicx}
\usepackage{amsmath}
\usepackage{amsthm}
\usepackage{eucal}
\usepackage{amssymb}
\usepackage{mathrsfs}
\usepackage{enumerate}
\usepackage{xcolor}

\newcommand{\Eref}[1]{Eq.~(\ref{#1})}

\newcommand{\eref}[1]{(\ref{#1})}

\newcommand{\Fref}[1]{Fig. \ref{#1}}

\newcommand{\fref}[1]{Fig. \ref{#1}}

\newcommand{\Sref}[1]{Section \ref{#1}}

\newcommand{\sref}[1]{section \ref{#1}}

\newcommand{\Aref}[1]{Appendix \ref{#1}}

\begin{document}

\title{Fast-mode elimination in stochastic metapopulation models}

\author{George W.~A.~Constable and Alan J.~McKane}

\affiliation{Theoretical Physics Division, School of Physics and Astronomy,
The University of Manchester, Manchester M13 9PL, United Kingdom}

\begin{abstract}
We investigate the stochastic dynamics of entities which are confined to a set of islands, between which they migrate. They are assumed to be one of two types, and in addition to migration, they also reproduce and die. Birth and death events are later moderated by weak selection. Systems which fall into this class are common in biology and social science, occurring in ecology, population genetics, epidemiology, biochemistry, linguistics, opinion dynamics, and other areas. In all these cases the governing equations are intractable, consisting as they do of multidimensional Fokker-Planck equations or, equivalently, coupled nonlinear stochastic differential equations with multiplicative noise. We develop a methodology which exploits a separation in time scales between fast and slow variables to reduce these equations so that they resemble those for a single island, which are amenable to analysis. The technique is generally applicable, but we choose to discuss it in the context of population genetics, in part because of the extra features that appear due to selection. The idea behind the method is simple, its application systematic, and the results in very good agreement with simulations of the full model for a range of parameter values.  
\end{abstract}
\maketitle

\section{Introduction}\label{secIntroduction}

Physicists now routinely apply the ideas and techniques of non-equilibrium statistical mechanics to areas outside of their own discipline. In the biological sciences, social sciences and in economics there are many instances of the interaction of the basic constituents of the systems giving rise to structures at the mesoscale or macroscale which can be studied and understood using the methodology of statistical physics~\cite{black2012,barton2011,castellano2009}. There are some crucial differences, however. In physics, atoms, molecules or spins frequently lie on regular grids or lattices and have an interaction strength depending on the distance between them. By contrast, conventional spatial modelling of this type will typically not be appropriate in a biological or social context. In these disciplines the interaction between individuals is more likely to be specified by a network, with links being of varying strengths or absent altogether~\cite{mendes2003,newman2010}. An alternative structure which is perhaps even more common is the situation where the nodes of the network are not individuals, but a population of individuals. The interaction is then between one population and another population connected to it by a link. The whole system is then a `population of populations' or metapopulation~\cite{levins1969}.

Metapopulations occur when modelling a wide variety of phenomena. In ecology, they may consist of disjoint pieces of vegetation able to support insects or animals, which are able to move from one area of vegetation to another~\cite{hanski1999}. In population genetics, they may be islands or `demes' of individuals with one of two different alleles of a particular gene, which are able to migrate from one deme to another~\cite{hartl1989}. In epidemiology, they may be cities with people who are infected or susceptible to a disease moving from one city to another~\cite{rozhnova2011}. In reaction kinetics, they may be compartments in which chemical reactions take place~\cite{challenger2013}. In linguistics the nodes may be people, who have two different linguemes --- two different ways of saying the same thing --- which they exchange with other people through conversation at a frequency given by the strength of the link~\cite{croft2000}. Similarly, in opinion dynamics, people may hold tokens which represent two different opinions, which may be exchanged with others~\cite{castellano2009}. The common mathematical structure in all these cases is a set of $\mathcal{D}$ `islands' which we will label by an index $i=1,\ldots,\mathcal{D}$. Island $i$ will contain a fixed number of individuals $N_i$, which undergo the processes of birth, death and migration from one island to another, in addition to perhaps other processes depending on the application.

A systematic analytic approach to investigating systems such as these begins at the microscale, that is, with the fundamental constituents of the system. It is assumed that the basic processes of birth, death, migration, etc.~can be described by a continuous time Markov process. The dynamics is then governed by a master equation~\cite{vankampen2007}, which, for any case of interest, will be intractable. It can be made less so by using a diffusion approximation in which the variables describing the state of the system are no longer the (discrete) number of individuals of a given type on island $i$, but the (continuous) fraction of that type of individual on island $i$. This approximation, valid for large $N_i$, turns the master equation into a Fokker-Planck equation (FPE)~\cite{mckaneBMB}. Although somewhat more tractable than the master equation, the FPE is still a partial differential equation in many variables, with a diffusion term which is state-dependent. It is equivalent to a set of stochastic differential equations (SDEs) with multiplicative noise~\cite{risken1989}, which is equally difficult to analyze.

There is a way forward in the analysis of equations such as these, providing that a set of `fast' modes can be identified, and eliminated systematically. This procedure, if it can be carried out, leaves a simpler set of equations which hopefully are amenable to analysis. We have recently described such a method, which was applied directly to a set of SDEs \cite{constable2013}, although there is a long history of techniques based on the same idea being applied to deterministic equations or to FPEs (for references see \cite{constable2013}).

In this paper we show that a variant of the procedure described in \cite{constable2013}, is in many instances ideally suited to reduce the population genetics of $\mathcal{D}$ demes to those of a single deme, with an effective population size and new effective parameters. This will be true even with variable deme size ($N_i \neq N_j$), as long as the sizes are of the same order, and also if selection is present. The technique is simple to understand and systematic --- effectively an algorithm --- with final expressions that are given as formulas involving, for example, eigenvalues and eigenvectors of matrices defined by the network.  While there has been work constructing reduced models of similar systems to the one we discuss here \cite{nagylaki1980SM,mckaneConsensus}, both the models and the techniques employed to reduce them, are less general than those we present. We also mention two approaches to the study of different systems which use some of the same ideas that we utilize~\cite{rogers2013,doering}.

The outline of the paper is as follows. In \Sref{secModel} we define the microscopic model for a set of $\mathcal{D}$ demes each containing individuals with two different alleles, and connected in a general network, and give the form of the corresponding SDEs. Initially the model is introduced for the selectively neutral case, in which neither of the alleles has an advantage over the other. The model is then naturally extended to the case where a selective bias exists. The details of the reduction method, whereby the $\mathcal{D}$ SDEs are reduced to a single SDE, are given in \Sref{secProjection}, along with formulas for the parameters appearing in the reduced equations. For clarity the method is presented in two broad regimes; the selectively neutral case, \Sref{secNeutral},  and the case with selective bias, \sref{secSelectionMigration}. In \Sref{secResults} we compare the probability of fixation of one of the alleles and the mean time to fixation calculated from the reduced SDEs with the results from simulating the original microscopic model for a range of parameters. The agreement is found to be very good. In \Sref{secLimits}, we conduct an investigation into the range of validity of the method. We find it continues to provide good predictions across a larger parameter range than one might na\"\i vely expect. In \Sref{secConclusion} we conclude and describe possible future avenues of investigation. A technical appendix gives details of the derivation of analytic results used in the main text.

\section{Model}\label{secModel}

As outlined in the Introduction, the system consists of a set of $\mathcal{D}$ islands or demes. On each of these demes is a fixed population of well-mixed individuals. The population can vary from deme to deme such that the population on the $i^{\mathrm{th}}$ deme is denoted by the integer $N_{i}=\beta_{i}N$, where $\beta_{i}$ is of order one and acts to moderate some typical deme size $N$. We assume that on each island there exist two types of haploid (single allele) organisms, carrying alleles $A$ and $B$ respectively. The number of $A$ alleles on the $i^{\mathrm{th}}$ island is denoted $n_{i}$. Since the population of each island is fixed this leaves $\beta_{i}N - n_{i}$ organisms carrying allele $B$. Having described the initial set-up we can proceed to consider the dynamical behaviour. 

The key aspect of the dynamics comes from the finite size of the population. Only in this case is there genetic drift, which is a consequence of the demographic stochasticity, or noise, generated from the finite and discrete nature of the population. In order to correctly and consistently include demographic stochasticity, it is necessary to begin from an individual based model (IBM). Rather than arrive at an equation which describes the definite state of the system at some time (a deterministic description), we require an equation which describes the probability that the system occupies any of the possible states at each time. To do this we first need to define the states of the system, which we assume are given by a vector of non-negative integers, $\bm{n}$, which in our case has as the $i^{\mathrm{th}}$ entry the number of individuals with allele $A$ on island $i$. We then need to specify a set of probability transition rates, $T(\bm{n'}|\bm{n})$. These give the probability per unit time that the system transitions from its current state, given by the vector $\bm{n}$, to a new state, $\bm{n'}$. Assuming that these transition rates only depend on the current state of the system, and not on its past history, the statistical time evolution of the system is then completely described by the master equation~\cite{vankampen2007}
\begin{equation}
\frac{d p(\bm{n},t)}{d t} = \sum_{\bm{n}'\neq\bm{n}}\,\left[ T(\bm{n}|\bm{n}')
p(\bm{n}',t) -  T(\bm{n}'|\bm{n})p(\bm{n},t) \right].
\label{master}
\end{equation}

\begin{figure}
\includegraphics[width=0.4 \textwidth]{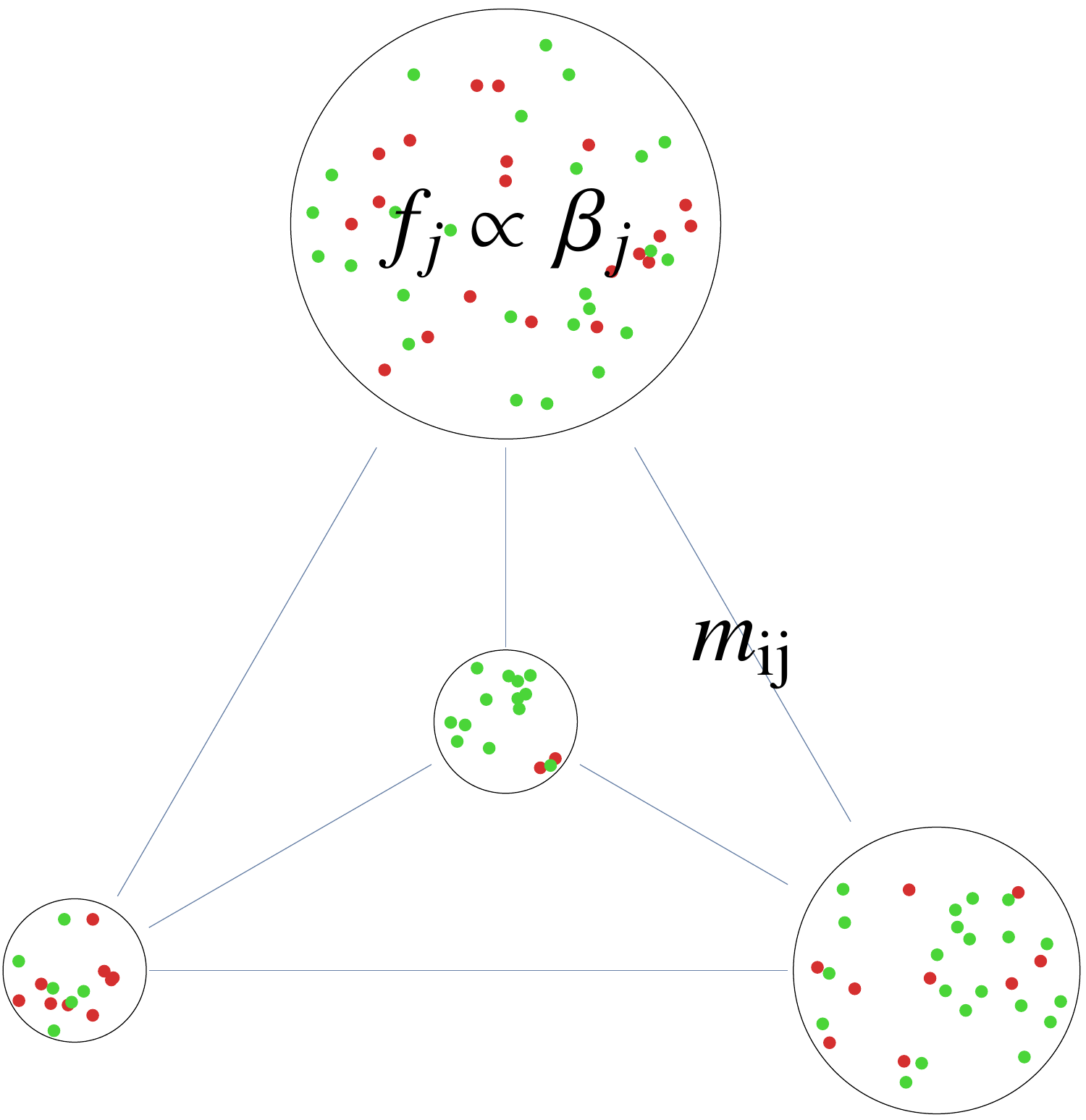}
\caption{(Color online) A simple depiction of the model: a set of $\mathcal{D}$ islands, each containing two types of individual, is connected by links whose strength is given by a product of the birth rates, $f_j$, and the migration probabilities $m_{ij}$.}
\label{simpleDiagram}
\end{figure}

The choice of the transition rates $T(\bm{n'}|\bm{n})$ fully define the dynamics. To construct these, we start by choosing an island, $j$, on which a reproduction event should take place based on a probability $f_{j}$ such that $\sum_{j=1}^{\mathcal{D}}f_{j}=1$. If the per capita birth rate on each island is equal, then $f_{j}$ will simply be proportional to the number of individuals on island $j$, so that we have $f_{j} = \beta_{j}/\sum_{k=1}^{\mathcal{D}}\beta_{k}$. Once an island has been chosen, an organism on the island must be picked to reproduce. If alleles $A$ and $B$ are selectively \emph{neutral} this probability will be entirely dependent on the local allele frequencies. Once an allele is picked to reproduce, the offspring must displace a pre-existing individual in one of the demes to keep the population of the demes fixed. The deme it chooses to migrate to is dependent on the migration matrix $m_{ij}$, the probability an offspring with progenitor in deme $j$ will migrate to deme $i$, normalized such that $\sum_{i}^{\mathcal{D}}m_{ij}=1$. The probability of offspring \emph{not} migrating can then be alternatively written $m_{ii} = 1 - \sum_{j \neq i}^{\mathcal{D}} m_{ji}$. Finally, the type of organism it replaces is proportional to the allele frequencies in the destination deme, $i$. This is essentially the Moran model~\cite{moran1957,moran1962}, with migration, depicted in \Fref{simpleDiagram}. A simple consideration of the combinatorics of this process leads to the transition rates,
\begin{eqnarray}\label{neutralMigTransitionRates}
 T( n_{i}+1 | n_{i} ) &=& \sum_{j=1}^{\mathcal{D}} \left( f_{j} \right) \left(\frac{n_{j}}{\beta_{j} N} \right) \left( m_{ij} \right) \left(\frac{\beta_i N - n_{i}}{\beta_{i}N - \delta_{ij}} \right)  \, ,
\nonumber \\			
 T(n_{i}-1 | n_{i} ) &=& \sum_{j=1}^{\mathcal{D}} \left( f_{j} \right) \left(\frac{\beta_{j}N-n_{j}}{\beta_{j}N} \right) \left( m_{ij} \right) \left(\frac{n_{i}}{\beta_{i}N - \delta_{ij}} \right) \, , \nonumber \\
\end{eqnarray}
where the dependence of the probability transition rates, $T(\bm{n'}|\bm{n})$, on elements of $\bm{n}$ that do no change in the transition has been suppressed. With all island sizes equal ($\beta_{i}=1\,\forall\,i$), this reduces to the migration model first discussed using this formalism in \cite{mckaneModels2007}. Since the only transitions are those where $n_{i}'=n_{i}\pm 1$, the master equation (\ref{master}) takes the much simpler form
\begin{align}
\frac{d p(\bm{n},t)}{d t} &= \sum_{i=1}^{\mathcal{D}} \left[ \vphantom{T(n_{i}+1|n_{i})p(n_{i},t)} T(n_{i}|n_{i}-1)p(n_{i}-1,t) \right. \nonumber \\
&\left. \vphantom{T(n_{i}|n_{i}-1)p(n_{i}-1,t)} -  
T(n_{i}+1|n_{i})p(n_{i},t) \right] \nonumber \\
& +  \sum_{i=1}^{\mathcal{D}} \left[ \vphantom{T(n_{i}-1|n_{i})p(n_{i},t)} 
T(n_{i}|n_{i}+1)p(n_{i}+1,t) \right. \nonumber \\
&\left. \vphantom{T(n_{i}|n_{i}+1)p(n_{i}+1,t)} - 
T(n_{i}-1|n_{i})p(n_{i},t) \right]\,.
\label{Meqn}
\end{align}

%%%%%%%%%%%%%%%%%%%%%%%%%%%%%%%%%%%%%%%%%%%%%%%%%%%%%%%%%%%%%%%%%%%%%%%%%%
%%%%%%%%%%%%%%%%%%%%%%%%%%%%%%%%%%%%%%%%%%%%%%%%%%%%%%%%%%%%%%%%%%%%%%%%%%%

%%%%%%%%%%%%%%%%%%%%%%%%%%%%%%%%%%%%%%%%%%%%%%%%%%%%%%%%%%%%%%%%%%%%%%%%%%
%%%%%%%%%%%%%%%%%%%%%%%%%%%%%%%%%%%%%%%%%%%%%%%%%%%%%%%%%%%%%%%%%%%%%%%%%%%

The situation becomes only slightly more complicated if a selective bias for one of the alleles exists. In this case, when picking which allele will reproduce, we must weight the local deme frequencies by the relative fitness of each allele. For simplicity, we consider here the case of frequency independent selection and weight the probability of selecting an $A$ allele on island $j$ by $1 + s \alpha_{j}$ and the $B$ allele by $1$. Here $s$ indicates the typical strength of selection, while $\alpha_{j}$ is a parameter that allows us to moderate this selection strength from island to island. The parameter $\alpha_{j}$ is positive if allele $A$ is advantageous in deme $j$, or negative if it is deleterious. The probability of picking $A$ and $B$ conditional on island $j$ first being chosen for reproduction are then respectively
\begin{align*}
 \frac{ (1 + s \alpha_{j} )n_{j}}{(1 + s \alpha_{j})n_{j}+  (\beta_{j}N-n_{j})} \ \ \textrm{and} \ \ \frac{ \beta_{j}N-n_{j} }{(1 + s \alpha_{j})n_{j}+  (\beta_{j}N-n_{j})}.
\end{align*}
Typically, the parameter $s$ will be small in biological situations, which allows us to Taylor expand in $s$ to arrive at the set of transition rates
\begin{eqnarray}
&&T( n_{i}+1 |  n_{i}) = \sum_{j= 1}^{\mathcal{D}} \left( f_{j} \right) \left( \frac{\beta_i N-n_{i}}{\beta_{i}N - \delta_{ij}} \right)  \left(m_{ij} \right) \times    \left( \frac{n_{j}}{\beta_{j}N} \right. \nonumber \\
		      && \left. + s \alpha_{j} \frac{n_{j}(\beta_{j}N-n_{j})}{(\beta_{j}N)^{2}} - s^{2}\alpha_{j}^{2} \frac{n_{j}^{2}(\beta_{j}N - n_{j})}{(\beta_{j}N)^{3}} + \mathcal{O}(s^{3}) \right) \, , \nonumber \\
\label{splusMigTransitionRates}
\end{eqnarray}
and
\begin{eqnarray}     
&&T(n_{i}-1 | n_{i} ) =\sum_{j=1}^{\mathcal{D}}  \left( f_{j} \right) \left( \frac{n_{i}}{\beta_{i}N-\delta_{ij}} \right) \left( m_{ij} \right) \times \left( \vphantom{s^{2}\alpha_{j}^{2} \frac{n_{j}^{2}(\beta_{j}N - n_{j})}{(\beta_{j}N)^{3}}} 1 - \frac{n_{j}}{\beta_{j}N} \right. \nonumber \\
                             &&\left.   - s \alpha_{j} \frac{n_{j}(\beta_{j}N-n_{j})}{(\beta_{j}N)^{2}} + s^{2}\alpha_{j}^{2} \frac{n_{j}^{2}(\beta_{j}N - n_{j})}{(\beta_{j}N)^{3}} + \mathcal{O}(s^{3}) \right) \, , \nonumber \\
\label{sminusMigTransitionRates}
\end{eqnarray}
for the generalized model with selection. We note that by setting $s=0$, we once again obtain the neutral model described by \Eref{neutralMigTransitionRates}.

While the master equations in both these cases are intractable, there exist a range of standard approximation techniques which can be used to simplify it. Historically, the one which has been used most extensively in population genetics is the diffusion approximation~\cite{fisher1922,crow2009} in which the variables $x_{i}=n_{i}/\beta_{i}N$ are introduced, and assumed to be continuous. This is clearly a large $N$ approximation, which allows the master equation to be Taylor expanded in $N^{-1}$, and then truncated at second order to obtain a generalized diffusion, or Fokker-Planck, equation. More systematically, we are starting from a Kramers-Moyal expansion~\cite{risken1989,gardiner2009}, and then using the master equation to calculate the jump-moments, showing that the third and higher moments are down by powers of $N^{-1}$~\cite{mckaneBMB}. This will be described in more detail elsewhere for the specific model we are considering in this paper~\cite{projectionBio}.

The FPE for the case of the transitions rates given by Eqs.~(\ref{splusMigTransitionRates}) and (\ref{sminusMigTransitionRates}) is 
\begin{align}\label{generalFPE}
\frac{\partial p(\bm{x},t)}{\partial t} = - \frac{1}{N} \sum_{i=1}^{\mathcal{D}} \frac{\partial}{\partial x_{i}} \left[A_{i}(\bm{x})p(\bm{x},t)\right]   \nonumber \\
                                                                               +    \frac{1}{2N^{2}}\sum_{i=1}^{\mathcal{D}}\frac{\partial^{2}}{\partial x_{i}^{2}} \left[B_{ii}(\bm{x})p(\bm{x},t)\right],
\end{align}
where the drift terms $A_{i}(\bm{x})$ and the diffusion terms $B_{ii}(\bm{x})$ are given below. A quantity which naturally appears in both $A_{i}(\bm{x})$ and $B(\bm{x})$ is $G$, a generalized migration rate matrix, which is the product of the birth rate, $f_{j}$, and the migration probability, $m_{ij}$:
\begin{eqnarray}
 G_{ij} = m_{ij}f_{j}, \quad \mathrm{where} \quad \sum_{i,j=1}^{\mathcal{D}}G_{ij} = 1;
\label{normalization_G}
\end{eqnarray}
the normalization of $G$ being inherited from the normalization of $m$ and $\bm{f}$. For the neutral case the drift vector is found to be
\begin{eqnarray}\label{defineNeutralA}
 A_{i}(\bm{x}) =  \frac{1}{\beta_{i}} \left(- x_{i}\sum_{j \neq i}^{\mathcal{D}}G_{ij} + \sum_{j \neq i}^{\mathcal{D}}G_{ij}x_{j} \right)\, ,
\end{eqnarray}
and the diffusion matrix is found to be diagonal, with elements
\begin{eqnarray}\label{generalFitnessDiffusion}
 B_{ii}(\bm{x})=   \frac{1}{\beta_{i}^{2}}\left(x_{i}\sum_{j=1}^{\mathcal{D}}G_{ij} + \sum_{j=1}^{\mathcal{D}}G_{ij}x_{j} \right. \nonumber \\ \left. - 2x_{i}\sum_{j=1}^{\mathcal{D}}G_{ij}x_{j} \right) \,.
\end{eqnarray}

In order to highlight the linearity of $A_{i}(\bm{x})$, it may be alternatively expressed as
\begin{eqnarray}
 A_{i}(\bm{x}) = \sum_{j=1}^{\mathcal{D}} H_{ij} x_{j},
\end{eqnarray}
where the matrix $H$ is defined by
\begin{equation}\label{defineH}
 H_{ij}=\frac{G_{ij}}{\beta_{i}} \quad  \, i \neq j, \qquad H_{ii} = -\sum_{j \neq i}^{\mathcal{D}} \frac{G_{ij}}{\beta_{i}} \, .
\end{equation}

In the case where selection is present, $s \neq 0$, we again obtain the FPE \eref{generalFPE}, but now with a drift term given by
\begin{flalign}\label{generalFitnessDrift}
 A_{i}(\bm{x}) &= \frac{1}{\beta_{i}}\left \{ \vphantom{\sum_{j=1}^{\mathcal{D}} G_{ij}\alpha_{j}^{2}x_{j}^{2}(1-x_{j}) } \sum_{j\neq i}^{\mathcal{D}} G_{ij}(x_{j} - x_{i}) + s \sum_{j=1}^{\mathcal{D}} G_{ij}\alpha_{j}x_{j}(1-x_{j})  \right. \nonumber \\
                       &\left. - s^{2} \sum_{j=1}^{\mathcal{D}} G_{ij}\alpha_{j}^{2}x_{j}^{2}(1-x_{j})  \right \} + \mathcal{O}(s^{3}) \, .
\end{flalign}
The diffusion matrix meanwhile is unchanged from the neutral case to leading order in $s$, \Eref{generalFitnessDiffusion}.

The Taylor expansion in $s$ is independent of the implementation of the Kramers-Moyal expansion. Since the Kramers-Moyal expansion involves a truncation at order $N^{-2}$, the appropriate truncation of the series in $s$ is ultimately dependent on the relative size of $s$ and $N$. We will find in the following analysis that it is sufficient for our purposes to work to the orders in $s$ indicated in Eqs.~(\ref{generalFitnessDrift}) and (\ref{generalFitnessDiffusion}).

\section{Projection method}\label{secProjection}

This simplified system, described by the approximately continuous state variable $\bm{x}$, is not much more analytically tractable than the original master equation. In addition, writing the dynamics as a partial differential equation does not aid the use of physical intuition in suggesting further simplifications. However, as is well-known~\cite{gardiner2009,risken1989}, the FPE~\eref{generalFPE}, is entirely equivalent to the It\={o} stochastic differential equation
\begin{equation}\label{generalSDE}
 \dot{x}_{i} = A_{i}(\bm{x}) + \frac{1}{\sqrt{N}}\eta_{i}(\tau),
\end{equation}
where the dot indicates differentiation with respect to $\tau= t/N$ and $\bm{\eta}(\tau)$ is a Gaussian correlated white noise with correlation functions
\begin{eqnarray}
\langle \eta_{i}(\tau)\eta_{j}(\tau') \rangle = B_{ij}(\bm{x}) \delta(\tau-\tau'). 
\end{eqnarray}
It is in this framework that we choose to work, applying concepts from deterministic dynamical systems theory in order to understand the stochastic problem, and introduce a method of simplification which has not been used before.

\subsection{The neutral case}\label{secNeutral}

Let us begin by considering the neutral case. The system is then governed by the SDE
\begin{equation}\label{linearNeutralSDE}
 \dot{x}_{i} = \sum_{j=1}^{\mathcal{D}}\,H_{ij} x_{j} 
+ \frac{1}{\sqrt{N}}\eta_{i}(\tau), \ \ \ i=1,\ldots,\mathcal{D}.
\end{equation}

First, we examine this equation in the deterministic limit, $N \rightarrow \infty$. The SDE \eref{linearNeutralSDE} is then solvable through a linear analysis. It is clear from the definition of the matrix $H$ in \Eref{defineH} that each row in $H$ sums exactly to $0$ for any choice of parameters. We may therefore write $\sum_{j=1}^{\mathcal{D}}\,H_{ij} = 0$ for all $i$, or alternatively as the eigenvalue equation $\sum_{j=1}^{\mathcal{D}}\,H_{ij}\,v^{(1)}_j = 0$, where $v^{(1)}_j =1\,\forall j$ is a right-eigenvector of $H$ with eigenvalue zero. We shall denote this eigenvector as $\bm{v}^{(1)}=\bm{1}$, where $\bm{1}$ is the $\mathcal{D}$-dimensional vector 
\begin{align}\label{eigenvector1}
\bm{1} \equiv \left( \begin{array}{c} 1 \\ 1 \\ \vdots \\ 1 \end{array} \right) \, , \quad  \mathrm{so\ that } \quad H \bm{1} = 0 \, .
\end{align}
In addition, it can be shown that all other eigenvalues, $\lambda^{(2)}\ldots \lambda^{(\mathcal{D})}$ have a real part which is negative, under the condition that $H$ is irreducible. In terms of our physical system, this amounts to specifying that no subgroup of demes is isolated from any other. 

To prove this, one can transform $H$ into a stochastic matrix. Firstly we introduce a matrix $\tilde{H}$, such that $\tilde{H}_{ij}=\beta_{\rm min}H_{ij}/(\mathcal{D}-1)$, where $\beta_{\rm min}$ is the smallest element of $\bm{\beta}$. Every off-diagonal element of $\tilde{H}$ then lies in the interval $[0,1]$, while every diagonal element lies in the interval $[-1,0)$. We now form the matrix $S$ with entries $S_{ij}=\tilde{H}_{ij}+\delta_{ij}$; since all entries of this matrix are non-negative, and since each row sums to one, $S$ is a stochastic matrix~\cite{gantmacher1959,cox1965}. This implies that largest eigenvalue of $S$ is $1$, and the magnitude of all its other eigenvalues is less than one~\cite{gantmacher1959,cox1965}. Further, by construction, $S$ and $H$ share the same set of eigenvectors. We can use these properties to show that the largest eigenvalue of $H$ is zero, while all other eigenvalues have a negative real part.

The right- and left-eigenvectors corresponding to eigenvalue $\lambda^{(i)}$ will be denoted by $\bm{v}^{(i)}$ and $\bm{u}^{(i)}$ respectively. They are orthogonal to each other, and will be normalized so that
\begin{equation}
\sum^{\mathcal{D}}_{k=1}\,u^{(i)}_k v^{(j)}_k = \delta_{i j}.
\label{orthonormal}
\end{equation}
In the special case that $H$ is symmetric the left- and right-eigenvectors coincide and the eigenvalues are real.

Already this tells us a great deal about the system dynamics in the deterministic limit, since the general solution to these equations is
\begin{eqnarray}
  x_i(\tau)=\sum^{\mathcal{D}}_{j=1} c_j v^{(j)}_i e^{\lambda^{(j)} \tau} \,,
\label{generalLinearSolution}
\end{eqnarray}
where the $c_j$ are constants. Both $\dot{x}_{i}$ and $A_{i}(\bm{x})= \sum_{j}H_{ij}x_j$ have a similar form to \eref{generalLinearSolution}, which means that after some time:
\begin{enumerate}[(i)]
 \item The vectors $\bm{x}$, $\dot{\bm{x}}$ and $\bm{A}(\bm{x})$ are all in the direction $\bm{1}$.
 \item The vectors $\dot{\bm{x}}$ and $\bm{A}(\bm{x})$ are actually zero in this direction, since $\lambda^{(1)}=0$.
\end{enumerate}
In other words, the eigenvalues $\lambda^{(i)}$ for $i\geq 2$ control the initial transient dynamics, which decay exponentially along the eigenvectors $\bm{v}^{(i)}$, $i\geq 2$, to some point on the vector $\bm{v}^{(1)}$ where the system will stay indefinitely. In the terminology of dynamical systems $\bm{v}^{(1)}$ is  coincident with a center manifold~\cite{wiggins2003}, and we shall refer to the eigenvectors $\bm{v}^{(i)}$, $i\geq 2$ as the fast directions.

Now, suppose we wish to ignore the initial fast behaviour of the system and pick out only the long-term dynamics. The condition that $\bm{A}(\bm{x})$ has no components in the fast directions $\bm{v}^{(j)}$, $j=2,\ldots,\mathcal{D}$ may be written in the form
\begin{equation}\label{centerManifold}
\sum_{i=1}^{\mathcal{D}} u^{(j)}_{i} A_{i}(\bm{x}) = 0 \, , 
\qquad j = 2, \ldots, \mathcal{D},
\end{equation}
which we will also make use of when discussing the non-neutral case. Since the state of the system, $\bm{x}$, lies on the line $\bm{1}$, we have $x_{1}=x_{2}=\ldots=x_{\mathcal{D}}$. We will denote the coordinate along $\bm{1}$ as $z$, so that the center manifold in the neutral case is simply
\begin{equation}
 x_{i} = z, \ \ i=1,\ldots,\mathcal{D}.
\label{define_center}
\end{equation}
However both $\dot{\bm{x}}$ and $\bm{A}(\bm{x})$ are zero on the center manifold, and so the value of $z$ does not change with time. Although the deterministic dynamics of the neutral model is trivial, the methodology developed here will be applicable to the case with selection, which has non-trivial dynamics.

\begin{figure}
 \includegraphics[width=0.45\textwidth]{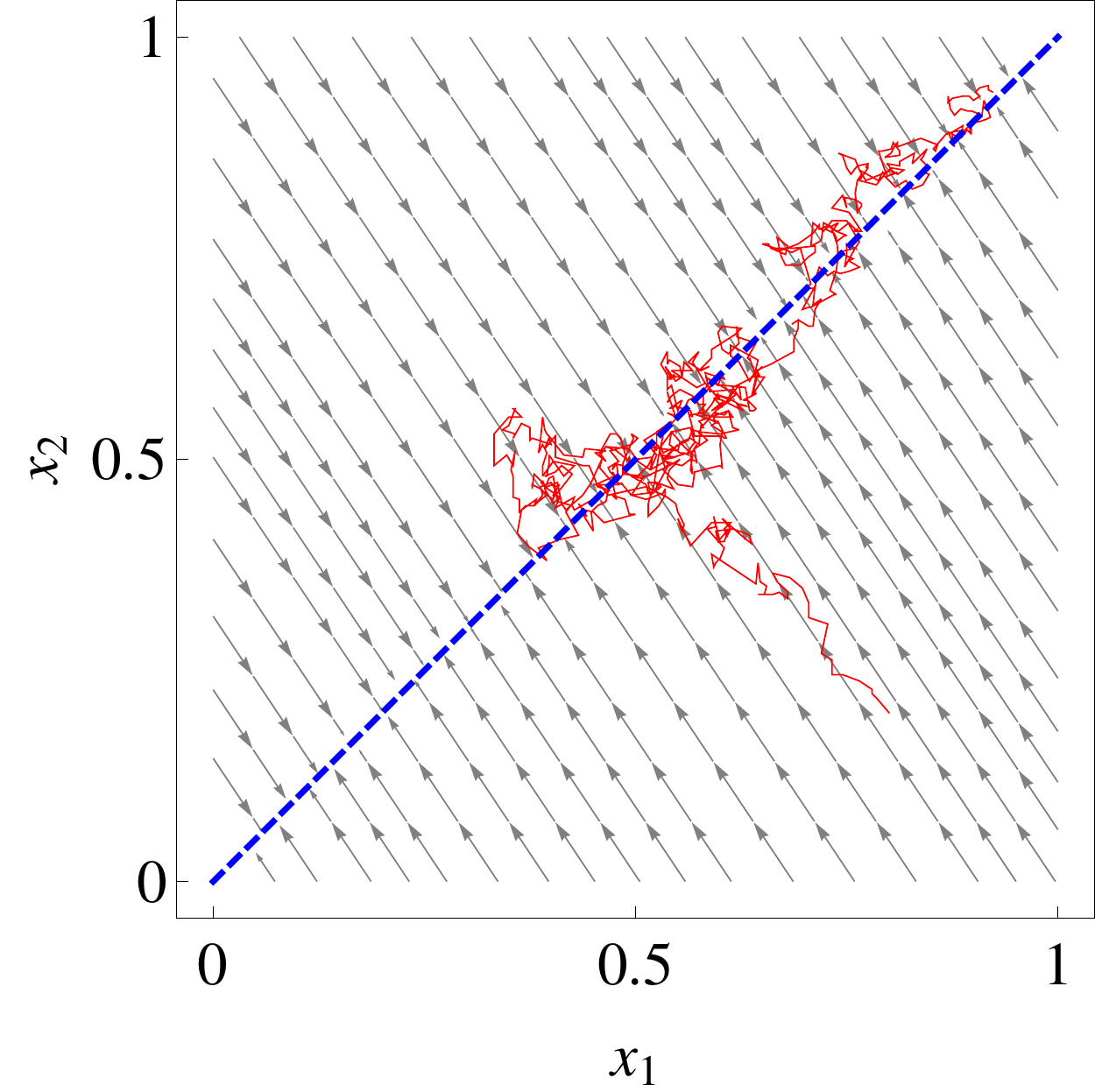}
 \includegraphics[width=0.45\textwidth]{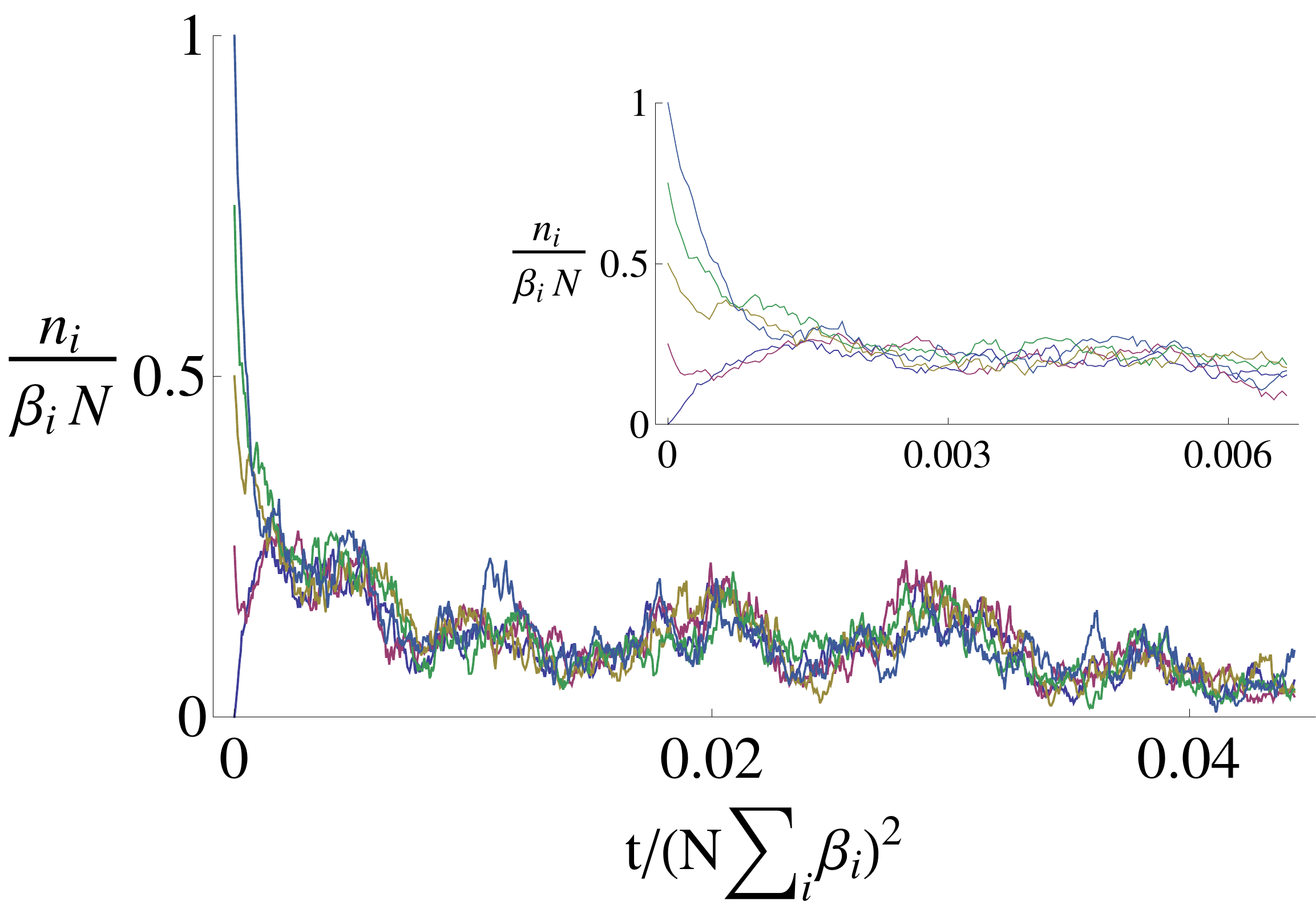}
\caption{(Color online) Upper panel: Time series of an individual stochastic trajectory (red) and deterministic trajectories (gray) for a neutral model with $\mathcal{D}=2$. The stochastic trajectory can be seen to quickly collapse onto the deterministic center manifold $x_{1}=x_{2}$, highlighted in blue, along which stochastic dynamics are observed. Lower panel: Time series for the populations in a $\mathcal{D}=5$ system with equal deme sizes. The inset graph shows clearly that the trajectories collapse to the center manifold, after which they are coupled and move in a stochastic fashion. The system size is $N=300$ for each deme in both plots.}
\label{neutralCollapse}
\end{figure}

We now ask, what happens when the population is finite and the stochastic dynamics play a role? We would expect that far from the center manifold, the deterministic dynamics would dominate over the noise terms, and drag the system to the center manifold, along which the stochastic dynamics would dominate. In turn any fluctuation that acted to move the system off the center manifold, would soon be quashed by the deterministic term. This is indeed what we see, as demonstrated in \fref{neutralCollapse} for two and five demes respectively. A clear separation of timescales exists; the deterministic dynamics act quickly to bring the system to the region of the center manifold, along which the system moves stochastically. Our intention is now to extend our treatment of the deterministic system, in which we sought to neglect initial transient dynamics, to the full stochastic system. 

%%%%%%%%%%%%%%%%%%%%%%%%%%%%%%%%%%%%%%%%%%%%%%%%%%%%%%%%%%%%%%%%%%%%%%%%%%
%%%%%%%%%%%%%%%%%%%%%%%%%%%%%%%%%%%%%%%%%%%%%%%%%%%%%%%%%%%%%%%%%%%%%%%%%%%

%%%%%%%%%%%%%%%%%%%%%%%%%%%%%%%%%%%%%%%%%%%%%%%%%%%%%%%%%%%%%%%%%%%%%%%%%%
%%%%%%%%%%%%%%%%%%%%%%%%%%%%%%%%%%%%%%%%%%%%%%%%%%%%%%%%%%%%%%%%%%%%%%%%%%%

With this in mind, we make the assumption that there is no noise in the fast directions. The only remaining contribution to the noise is then in the direction $\bm{1}$: $\bm{\eta}(\tau)= \zeta(\tau)\bm{1}$. Since, $\bm{x}$ is restricted to the center manifold, and there is no deterministic drift along it, the SDE (\ref{linearNeutralSDE}) simply reduces to 
\begin{equation}\label{neutralSDE1D}
\dot{z} = \frac{1}{\sqrt{N}}\zeta(\tau).
\end{equation}
The noise $\zeta$ can be characterized by using Eq.~(\ref{orthonormal}) to write it in terms of $\bm{\eta}$ as $\zeta(\tau)=\sum_{i}u^{(1)}_{i}\eta_{i}(\tau)$. Then we see that $\zeta$ is a Gaussian white noise with zero mean and correlator 
\begin{eqnarray}
\langle \zeta(\tau)\zeta(\tau') \rangle = \sum^{\mathcal{D}}_{i,j=1} u^{(1)}_{i}\left.B_{ij}(\bm{x})\right|_{\bm{x}=z\bm{1}} u^{(1)}_{j}\delta(\tau-\tau'), 
\end{eqnarray}
where $B_{ij}(\bm{x})$ has been evaluated on the center manifold $\bm{x}=z\bm{1}$. Setting $\bm{x}=z\bm{1}$ in Eq.~(\ref{generalFitnessDiffusion}) with $s=0$ we find
\begin{eqnarray}\label{defineBbar}
\bar{B}(z) &\equiv& \sum^{\mathcal{D}}_{i,j=1} u^{(1)}_{i}\left.B_{ij}(\bm{x})\right|_{\bm{x}=z\bm{1}} u^{(1)}_{j} \nonumber \\
&=&  2 z(1 - z)\,\sum_{i,k=1}^{\mathcal{D}}\,[u^{(1)}_{i}]^{2}G_{ik}\beta_{i}^{-2}
\nonumber \\
&\equiv& 2 b_{1} z (1 - z) \, ,
\end{eqnarray}
where we have introduced the constant
\begin{equation}
 b_{1} = \sum_{i,k=1}^{\mathcal{D}}\,[u^{(1)}_{i}]^{2}G_{ik}\beta_{i}^{-2}\, ,
\label{b_1}
\end{equation}
and use a bar to indicate evaluation on the center manifold. Extending this notation, \Eref{neutralSDE1D} may be more generally expressed as
\begin{eqnarray}\label{generalSDE1D}
 \dot{z} = \bar{A}(z) + \frac{1}{\sqrt{N}}\zeta(\tau),
\end{eqnarray}
where the drift term evaluated on the center manifold is zero, $\bar{A}(z)=0$, and where $\zeta(t)$ is Gaussian correlated white noise with zero mean and correlation function
\begin{eqnarray}
\langle \zeta(\tau) \zeta(\tau') \rangle =  \bar{B}(z)\delta(\tau-\tau'). 
\label{zeta_corr}
\end{eqnarray}

Our aim in this subsection has been to characterize the stochastic dynamics along the center manifold, and so to develop a one-dimensional, reduced theory. We have also observed that given some set of initial conditions, the stochastic system relaxes to the center manifold on a much faster timescale than that on which the stochastic dynamics act. The reduced model is therefore ideally suited to answering questions related to global fixation, such as the fixation probability and the mean fixation time, which are long-time properties of the system. 

In order to approximate the initial value of the system on the center manifold, we assume that the trajectory to the center manifold is essentially deterministic. Then the initial condition on the center manifold, $z_{0}$, is simply the component of the full initial condition $\bm{x}_{0}$, along $\bm{v}^{1}=\bm{1}$:
\begin{eqnarray}
 z_{0} = \sum_{i=1}^{\mathcal{D}}u^{(1)}_{i} x_{0i} \, .\label{initialConditon}
\end{eqnarray}
Together with Eqs.~({\ref{generalSDE1D}) and (\ref{zeta_corr}), this fully defines the reduced model.

\subsection{The case with selection}\label{secSelectionMigration}

Having considered the neutral model with migration, in which neither allele $A$ nor $B$ has any fitness relative to the other, we now consider the case where there is a relative fitness.

To reiterate, this model is now described by the SDE \eref{generalSDE}, with drift and diffusion terms given by \Eref{generalFitnessDrift} and \Eref{generalFitnessDiffusion} to leading order. We note that deterministically the system may now admit a non-trivial fixed point in the region between $0 < x_{i} < 1,\,i=1,\ldots,\mathcal{D}$; the analysis of the consequences of this will be discussed elsewhere~\cite{projectionBio}.

We begin by noting that since $s$ has been defined as a small parameter in relation to the migration matrix (and hence also the matrix $H$), we would still expect the system to exhibit a separation of timescales. Now however there is no center manifold; there is no line along which the deterministic dynamics vanish, as the nonlinear $s$ terms cause deterministic drift. Instead, a slow subspace exists onto which the system quickly relaxes. The existence of such a subspace is clearly seen in deterministic and stochastic simulations. We avoid the term ``slow manifold'', since this has a distinct technical meaning~\cite{berglund2006}; the slow subspace is a curved line in the present context, although because $s$ is small it only has a slight curvature.

How do we mathematically specify the slow subspace? Although Eq.~(\ref{generalFitnessDrift}) shows that the $s \neq 0$ deterministic theory is inherently nonlinear, since $s$ is typically very small, we will continue to use the $s=0$ left-eigenvectors $\bm{u}^{(j)}$ to approximate the slow subspace through Eq.~(\ref{centerManifold}). Solving these equations numerically, we find that they provide a very good approximation to the observed slow subspace. In Appendix \ref{appSlowManifold} we solve the $(\mathcal{D}-1)$ equations of Eq.~(\ref{centerManifold}) to find the slow subspace analytically. To do this we transform to a co-ordinate system $\bm{x} = z\bm{1} + \sum^{\mathcal{D}-1}_{a=1} y_{a} \bm{v}^{(a+1)}$. We find that the equation of the slow subspace takes the form
\begin{equation}
y_a(z) = c_{a} s z(1-z) + \mathcal{O}(s^2),
\label{slow_subspace}
\end{equation}
where the $c_a$ are constants which are calculated in the Appendix. As $s \to 0$, $y_a(z) \to 0$, and the slow subspace becomes the center manifold, $\bm{1}$, of the neutral case discussed in Section \ref{secNeutral}. 

The condition given by Eq.~(\ref{centerManifold}) still means that $\bm{A}$ lies in the $\bm{1}$ direction. Therefore once again the deterministic dynamics is only in this direction. If we ask that the noise is only in this direction too, then we again write $\bm{\eta}(\tau)= \zeta(\tau)\bm{1}$, just as we did in the neutral case, to find Eqs.~(\ref{generalSDE1D}) and (\ref{zeta_corr}), but now with $\bar{A}(z) \neq 0$. To find $\bar{A}(z)$ we use Eq.~(\ref{orthonormal}) to pick out the component of $\bm{A}$ along $\bm{1}$, and evaluate it in the slow subspace given by Eq.~(\ref{slow_subspace}):
\begin{equation}
\bar{A}(z) \equiv \sum^{\mathcal{D}}_{i=1} u^{(1)}_{i} A_{i}(z,\bm{y}(z))\,.
\label{A_bar}
\end{equation}
It is important to note that the dynamics is only in the direction $\bm{1}$, even though the system relaxes to the slow subspace given by Eq.~(\ref{slow_subspace}). Thus the approximation requires that the $y_a$ have no dynamics; they are simply mapped on to a value of $z$ using Eq.~(\ref{slow_subspace}).

In fact from a mathematical point of view, the whole procedure may be specified in terms of a projection onto the direction $\bm{1}$ from any point in state space, together with an understanding that the drift should be evaluated on Eq.~(\ref{slow_subspace}). To do this, we construct a matrix $P$, such that when it is applied to any vector wipes out the fast directions $\bm{v}^{(a)}$ for $a=2, \ldots , \mathcal{D}$, but leaves the component along the direction $\bm{v}^{(1)}= \bm{1}$,  untouched. Using the vector $\bm{u}^{(1)}$, which is perpendicular to the fast directions, we can construct the projection matrix as 
\begin{eqnarray}
 P_{ij} = \frac{v^{(1)}_{i}u^{(1)}_{j}}{\sum_{k=1}^{\mathcal{D}} v^{(1)}_{k} u^{(1)}_{k}}.
\end{eqnarray}
Since $\bm{v}^{(1)}$ is simply $\bm{1}$,
\begin{eqnarray}\label{defineProjection}
 P_{ij} =  u^{(1)}_{j}\, , \qquad  \, i = 1, \ldots , \mathcal{D},
\end{eqnarray}
using the orthonormality condition (\ref{orthonormal}). We have already implicitly used this to define $\bar{A}$ and $\bar{B}$, but it can be directly applied to Eq.~(\ref{generalSDE}) to obtain Eq.~(\ref{generalSDE1D}). The requirement that the drift vanishes in the fast directions, Eq.~(\ref{centerManifold}), is still required to find the slow subspace Eq.~(\ref{slow_subspace}).

Let us now apply this approximation procedure. We begin by noting that the drift vector \Eref{generalFitnessDrift} truncated at second order in $s$ can be alternatively expressed by
\begin{eqnarray}
 A_{i}(\bm{x}) &= \sum_{j=1}^{\mathcal{D}}\,H_{ij}x_{j} + \beta^{-1}_{i}\left \{ \vphantom{ \sum_{j=1}^{\mathcal{D}} G_{ij}\alpha_{j}^{2}x_{j}^{2}(1-x_{j}) } 
							    s \sum_{j=1}^{\mathcal{D}} G_{ij}\alpha_{j}x_{j}(1-x_{j}) \right. \nonumber \\ 
                       &\left. -s^{2} \sum_{j=1}^{\mathcal{D}} G_{ij}\alpha_{j}^{2}x_{j}^{2}(1-x_{j})  \right \} \, ,
\label{A_second_order}
\end{eqnarray} 
while the diffusion matrix is left unchanged from the neutral case to leading order (see \Eref{generalFitnessDiffusion}). In \Aref{appSlowManifold}, we obtain the expression \eref{generalFitnessDriftSlowManifold} for the drift vector evaluated on the slow subspace ($A_{i}(z,\bm{y}(z))$ in \Eref{A_bar}) in terms of the projected variable $z$. The elements of $\bm{A}(z)$ to this order take the form 
\begin{eqnarray*}
 A_{i}(z) = &-& s q^{(0)}_{i} z (1 - z) + s q^{(1)}_{i} z(1 - z)  - s^{2} q^{(2)}_{i} z^{2}(1 - z) \nonumber \\ &-& s^{2} q^{(3)}_{i} z (1 - z) (1 - 2 z)  \, ,
\end{eqnarray*}
where the vectors of parameters, $\bm{q}^{(i)}$, $i=0, \ldots, 3$, are defined in \Aref{appSlowManifold}. The diffusion matrix evaluated on the slow subspace is given in \Eref{generalFitnessDiffusionSlowManifold}.

As in \Sref{secNeutral} we apply the projection (\Eref{defineProjection}) to the SDE \eref{generalSDE}, with the above drift vector and diffusion matrix. This again leads to the reduced SDE of type \Eref{generalSDE1D}, but now with $\bar{A}(z)$ given by \Eref{A_bar}. The term $-s q^{(0)}_{i} z (1 - z)$ which appears in $A_i(z)$ does not appear in $\bar{A}(z)$ because $\sum_i u^{(1)}_{i}q^{(0)}_{i}=0$, which follows from $\sum_i u^{(1)}_{i}H_{ij}=0$. The remainder of the expression is given by
\begin{eqnarray}\label{ABarSNHalf}
 \bar{A}(z) = s a_{1}  z (1 - z) + s^{2} a_{2} z^{2}(1-z) \nonumber \\ + s^{2} a_{3} z(1-z)(1-2 z) \, ,
\end{eqnarray}
and $\bar{B}(z)$ retains the form obtained in the neutral case, \Eref{defineBbar}. The parameters $a_{1}$ and $a_{2}$ are found to be only dependent on the parameters of the problem ($m$, $\bm{f}$, $\bm{\beta}$, $\bm{\alpha}$) and on the left-eigenvector $\bm{u}^{(1)}$:
\begin{eqnarray}\label{definea1}
a_{1} = \sum_{i=1}^{\mathcal{D}} P_{ki}q^{(1)}_{i}  = \sum_{i,j=1}^{\mathcal{D}}u^{(1)}_{i} \frac{G_{ij}\alpha_j}{\beta_{i}}\,
\end{eqnarray}
and
\begin{eqnarray}
 a_{2} = - \sum_{i=1}^{\mathcal{D}} P_{ki}q^{(2)}_{i}  
= - \sum_{i,j=1}^{\mathcal{D}}u^{(1)}_{i} \frac{G_{ij}\alpha_j^2}{\beta_{i}}\,.
\end{eqnarray}
The parameter $a_{3}$ meanwhile is found to be dependent on the full set of left- and right-eigenvectors and their corresponding eigenvalues:
\begin{eqnarray}
& & a_{3} = - \sum_{i=1}^{\mathcal{D}} P_{ki}q^{(3)}_{i}  \label{definea3} \\
      &=& - \sum_{a=1}^{\mathcal{D}-1} \left[\sum_{i,j=1}^{\mathcal{D}}
\frac{u^{(1)}_{i}G_{ij}\alpha_j}{\beta_{i}} \sum_{k,l=1}^{\mathcal{D}} 
\frac{v^{(a+1)}_{j} u^{(a+1)}_{k}}{\lambda^{(a+1)}} 
\frac{G_{kl}\alpha_l}{\beta_{k}} \right]. \nonumber
\end{eqnarray}
Its more complicated form is a consequence of the curvature of the slow subspace.

\section{Comparison with simulations}\label{secResults}

In the last section, fast-variable elimination was used to simplify the migration model to a one-dimensional SDE with the form of \Eref{generalSDE1D}. The drift term $\bar{A}(z)$ is zero for the case $s=0$ and given by \Eref{ABarSNHalf} to second order in $s$. The diffusion term $\bar{B}(z)$ is given by \Eref{defineBbar} in each case.

Having started from an IBM, the validity of the approximation can now be tested by comparing predictions of the model against Gillespie simulations~\cite{gillespie1976,gillespie1977}, which provide exact solutions to realizations of the master equation \eref{Meqn}. As measures to test the validity of the approximation, we choose the fixation probability and time to fixation of the system given a set of initial allele frequencies on each deme.

First we note that the one-dimensional It\={o} SDE \eref{generalSDE1D} that we have constructed is equivalent to a one-dimensional FPE~\cite{gardiner2009,risken1989} which takes the form
\begin{eqnarray}\label{FPE1D}
 \frac{\partial p(z,t)}{\partial t} = &-& \frac{1}{N}  \frac{\partial}{\partial z} \left[\bar{A}(z)p(z,t)\right] \nonumber \\
&+& \frac{1}{2N^{2}}\frac{\partial^{2}}{\partial z^{2}} 
\left[\bar{B}(z)p(z,t)\right].
\end{eqnarray}
The fixation probability of allele $A$, $Q(z_{0})$, and time to fixation, $T(z_{0})$, as a function of the initial condition, projected onto the manifold
\begin{eqnarray}\label{projectedInitialConditions}
 P\bm{x}_{0} = z_{0} \bm{1} \, ,
\end{eqnarray}
can be calculated from the backward Fokker-Planck equation, formally the adjoint of the FPE~\cite{gardiner2009,risken1989}
\begin{eqnarray}\label{BFPE}
\frac{\partial q(z,t)}{\partial t} = \frac{\bar{A}(z)}{N}  \frac{\partial}
{\partial z} \left[q(z,t)\right] + 
\frac{\bar{B}(z)}{2N^{2}}\frac{\partial^{2}}{\partial z^{2}} \left[q(z,t)\right].
\end{eqnarray}
The general theory which starts from a FPE of the form given by Eq.~(\ref{FPE1D}), and via the backward equation (\ref{BFPE}) leads to equations for $Q(z_0)$ and $T(z_0)$, is standard and we refer the reader to the literature~\cite{gardiner2009,risken1989} for the details. The probability of fixation, $Q(z_0)$ satisfies the ordinary differential equation 
\begin{eqnarray}\label{generalODEFixProb}
 \frac{\bar{A}(z_0)}{N} \frac{dQ(z_{0})}{dz_{0}} + \frac{\bar{B}(z_{0})}{2N^{2}}\frac{d^{2}Q(z_{0})}{dz_{0}^{2}} = 0, 
\end{eqnarray}
with boundary conditions
\begin{eqnarray}
 Q(0)=0 \, , \qquad Q(1)=1 \, .
\end{eqnarray}
The boundary conditions may be understood by noting that when there are no instances of allele $A$ in the population, $z_{0}=0$, there is no possibility of fixation, while if there are no $B$ alleles, $z_{0}=1$ and allele $A$ has fixed.

The equation for the mean time to fixation, $T(z_{0})$, is
\begin{eqnarray}\label{generalODEFixTime}
\frac{\bar{A}(z_0)}{N} \frac{dT(z_{0})}{dz_{0}} + 
\frac{\bar{B}(z_{0})}{2N^{2}}\frac{d^{2}T(z_{0})}{dz_{0}^{2}} = -1,
\end{eqnarray}
with the boundary conditions,
\begin{eqnarray}
 T(0)=0 \, , \qquad T(1)=0 \,.
\end{eqnarray}
In this case the boundary conditions follow by noting that they are the time to fixation of \emph{either} $A$ or $B$ allele, so that at both extremes, $z_{0}=0$ and $z_{0}=1$, the system has already fixed on $A$ or $B$.

Solutions to these equations will now be described for the reduced dimension model in the cases $s=0$ and $s \neq 0$, and the predictions from the reduced systems compared to simulation.

\subsection{Comparison with simulations --- Neutral Case}

We begin by noting that the reduced Fokker-Planck equation now takes on the same functional form as that which is derived for the neutral Moran model on a single island \cite{crow2009}. Scaling $N^{2}$ by $b_{1}$ and setting $z=x_{1}$, the results are identical. 

For a general system with known parameters, the calculation of $b_{1}$ depends only on obtaining the left-eigenvector of $H$, $\bm{u}^{(1)}$. For small values of $\mathcal{D}$, or alternatively, $H$ matrices with some exploitable symmetries, it may be possible to obtain this analytically. Numerically however, expressions for $b_{1}$ are easily obtainable. One could thus proceed in an almost algorithmic way to obtain the reduced FPE given a migration matrix, $m$, island sizes, $\bm{\beta}$ and the island birth rates $\bm{f}$. 

We now note some special cases. Firstly, let us take the case where the matrix $H$ is symmetric. Then the left- and right-eigenvectors, $\bm{u}^{(1)}$ and $\bm{v}^{(1)}$ coincide (up to an overall constant). Since $\bm{v}^{(1)}=\bm{1}$, $u^{(1)}_i =$ constant for all $i$. Using the normalization condition (\ref{orthonormal}) we find that $u^{(1)}_i=\mathcal{D}^{-1}\ \forall i$, and so from Eq.~(\ref{b_1})
\begin{eqnarray}
b_1 = \sum_{i,j=1}^{\mathcal{D}} \frac{G_{ij}}{\left( \beta_{i}\mathcal{D} \right)^2}\, .
\end{eqnarray}

Perhaps more interesting is the case where the matrix $G$ is symmetric. We begin by looking at the quantity $\sum^{\mathcal{D}}_{i=1} \beta_{i}H_{ij}$ and expressing $H$ in terms of $G$ using Eq.~(\ref{defineH}):
\begin{eqnarray}
\sum^{\mathcal{D}}_{i=1} \beta_{i}H_{ij} &=&  
\sum_{i \neq j}^{\mathcal{D}} \beta_{i} \frac{G_{ij}}{\beta_i} + 
\beta_{j}\,\left( - \sum_{k \neq j}^{\mathcal{D}} \frac{G_{jk}}{\beta_{j}} \right) 
\nonumber \\
= \sum_{i \neq j}^{\mathcal{D}} G_{ij} &-& \sum_{k \neq j}^{\mathcal{D}} G_{jk} =
\sum_{i \neq j}^{\mathcal{D}} \left[ G_{ij} - G_{ji} \right] = 0,
\end{eqnarray}
if $G$ is symmetric. So in this case $\beta_i$ is the left-eigenvector of $H$ with zero eigenvalue, that is, it is equal to $u^{(1)}_i$, up to an overall constant. Since from Eq.~(\ref{orthonormal}), $\sum^{\mathcal{D}}_{i=1}u^{(1)}_i = 1$, we have that
\begin{eqnarray}
u^{(1)}_{i} = \frac{ \beta_{i} }{ \sum_{j=1}^{\mathcal{D}}\beta_{j} } \, .
\end{eqnarray}
This leads to 
\begin{eqnarray}
b_{1} = \left( \sum_{j=1}^{\mathcal{D}} \beta_{j} \right)^{-2} \, , \label{defineb1_symmetric}
\end{eqnarray}
using Eqs.~(\ref{normalization_G}) and (\ref{b_1}). This shows that, for a symmetric $G$ matrix, the reduced FPE for the metapopulation model is identical to the full FPE for a well-mixed model with the same total number of individuals, $N\sum_{i=1}^{\mathcal{D}}\beta_{i}$.

In the neutral case \Eref{generalODEFixProb} can be solved trivially to give
\begin{eqnarray}
 Q(z_{0}) = z_{0},\label{Q_s_0}
\end{eqnarray}
where we recall that the projected initial condition, $z_{0}$ is found from \Eref{projectedInitialConditions}. 
The probability of fixation depends only on the structure and form of the metapopulations through the determination of the initial conditions.

The mean time to fixation can be calculated from \Eref{generalODEFixTime}, and also resembles the standard result for one island~\cite{ewens2004}:
\begin{eqnarray}
T(z_{0})= - \frac{N^2}{b_{1}} \left[  (1-z_{0})\ln{(1 - z_{0})}  +  z_0 \ln{(z_0)} \right].\label{T_s_0}
\end{eqnarray}

In order to test the predictions of the reduced model, \Eref{Q_s_0} and \Eref{T_s_0}, we compare them against stochastic Gillespie simulations of the underlying microscopic model (specified by the transition rates \eref{neutralMigTransitionRates}) for a range of systems. We find excellent agreement. In particular, \Fref{Qs0} illustrates the results obtained from three different migration matrices, $m$. As predicted the probability of fixation, $Q(z_{0})$ is only dependent on the structure of $m$ though the projected initial condition, $z_{0}$ (see \Eref{initialConditon}). The mean time to fixation meanwhile is highly dependent on the structure of $m$. It is clear that the reduced model reflects this dependency very well.

%%%%%%%%%%%%%%%%%%%%%%%%%%%%%%%%%%%%%%%%%%%%%%%%%%%%%%%%%%%%%%%%%%%%%%%%%%
%%%%%%%%%%%%%%%%%%%%%%%%%%%%%%%%%%%%%%%%%%%%%%%%%%%%%%%%%%%%%%%%%%%%%%%%%%%

\begin{figure}
 \includegraphics[width=0.45\textwidth]{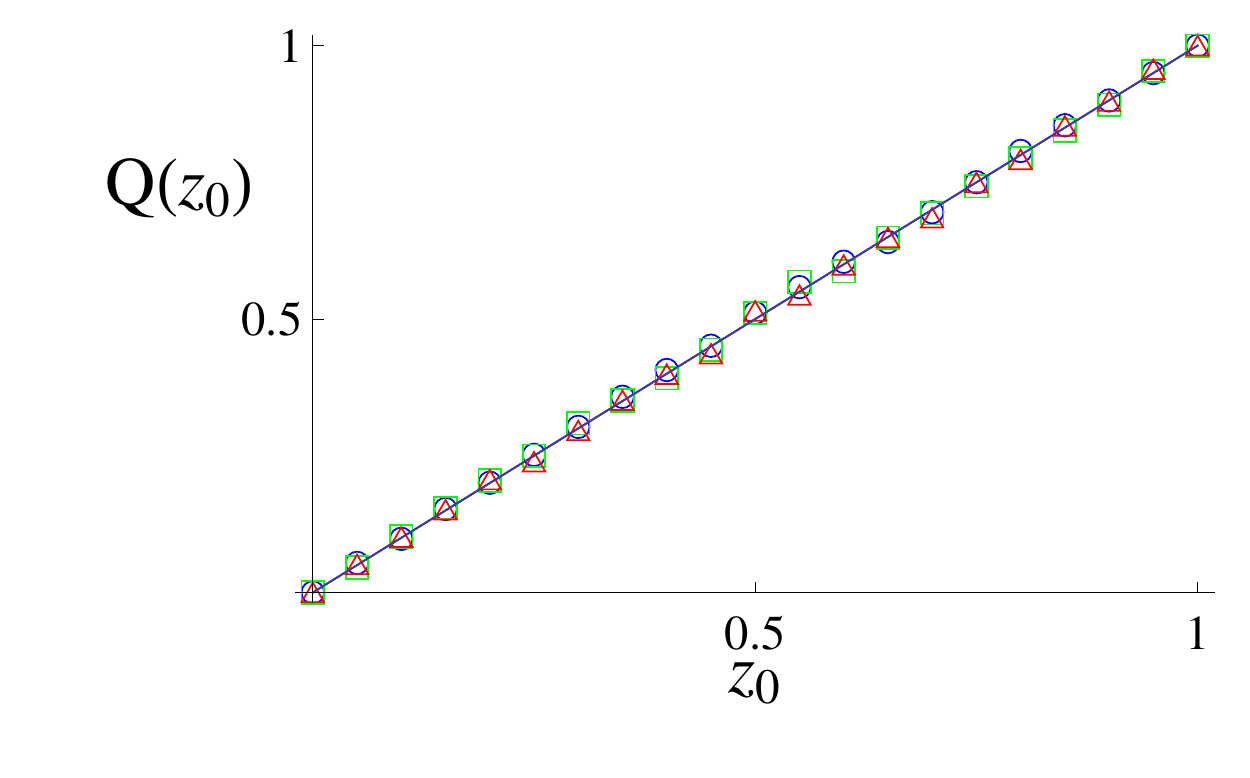}
 \includegraphics[width=0.45\textwidth]{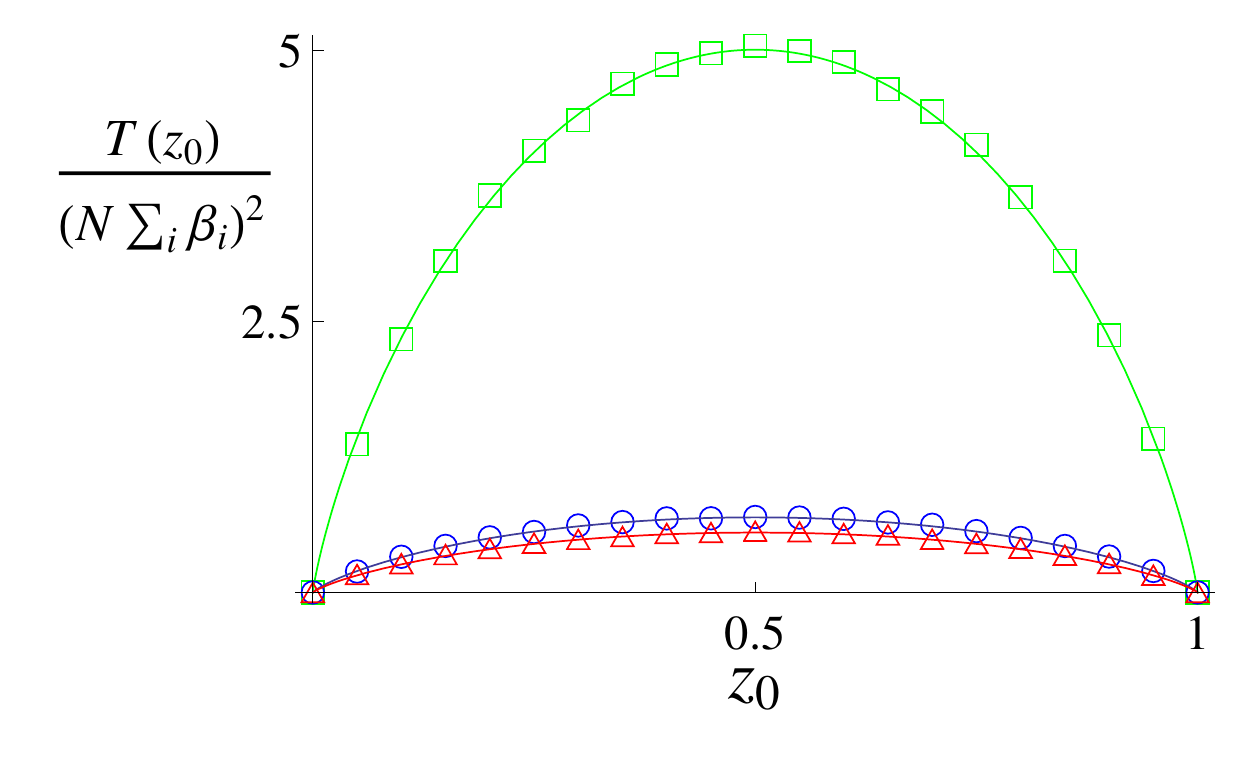}
\caption{(Color online) Upper panel: Probability of fixation as a function of the projected initial conditions for neutral systems, $s=0$. Lower panel: Mean time to fixation as a function of the projected initial conditions, again for $s=0$ neutral systems. Continuous lines show the analytic predictions from \Eref{Q_s_0} and \eref{T_s_0} while the values of symbols are obtained as the mean of $5000$ stochastic simulations. Here the model with three demes is studied, each of the demes has size $N=200$. The various colors/symbols are obtained from differing migration matrices. The system indicted by a blue line (the central data in the lower panel) has a symmetric migration matrix with $b_{1} = \left( \sum_{j=1}^{\mathcal{D}} \beta_{j} \right)^{-2}$.}
\label{Qs0}
\end{figure} 

%%%%%%%%%%%%%%%%%%%%%%%%%%%%%%%%%%%%%%%%%%%%%%%%%%%%%%%%%%%%%%%%%%%%%%%%%%
%%%%%%%%%%%%%%%%%%%%%%%%%%%%%%%%%%%%%%%%%%%%%%%%%%%%%%%%%%%%%%%%%%%%%%%%%%%

\subsection{Comparison with simulations --- case with selection}

In the following analysis we shall consider the case in which $\bar{A}(z)$ given by Eq.~(\ref{ABarSNHalf}) is truncated at first and second order in $s$ separately. There are two reasons for doing this. First, the case where only the linear term in $s$ is retained can again be mapped onto the full FPE for a one-island Moran model, but with selection. Now not only is $N^2$ scaled by $b_1$, but $s$ is scaled by $a_1$ and $\sqrt{b_{1}}$. Second, the order at which we truncate $\bar{A}(z)$ can be viewed as an assumption about the relative size of the parameters $s$ and $N$. In the Kramers-Moyal expansion, which was used to obtain \eref{generalFPE}, we have neglected terms smaller than order $N^{-2}$. For consistency we would also like to neglect any terms smaller than this arising from $s$ contributions. If  $s \approx \mathcal{O}(N^{-1})$, neglecting $\mathcal{O}(N^{-3})$ terms in \eref{generalFPE} results in an expression for $\bar{A}(\bm{x})$ which is first order in $s$. However, if $s \approx \mathcal{O}(N^{-1/2})$, neglecting $\mathcal{O}(N^{-5/2})$ terms results in $\bar{A}(x)$ which is second order in $s$. Finally, we will present results for rather small values of $N$ and rather large values of $s$, as compared with those commonly found in population genetics. We would expect our approximation scheme to become better as $N$ increases, and as a consequence of the above argument, give similar results for proportionally smaller values of $s$.

We begin by investigating the case where $\bar{A}(z)$ is truncated at first order in $s$. Solving Eq.~(\ref{generalODEFixTime}), with $\bar{B}(z)$ given by Eq.~(\ref{defineBbar}), leads to the probability of fixation being given by
\begin{eqnarray}
Q(z_{0}) = \frac{1 - \exp{\left(- N s a_{1} z_{0}/b_{1} \right)}}
{1 - \exp{\left(- N s a_{1}/b_{1} \right)}}.
\label{Q_first_order}
\end{eqnarray}
Scaling $N$ by a factor $a_1/b_1$ gives the one island result with $z=x_{1}$~\cite{ewens2004}. The form of $T(z_0)$ is found by solving Eq.~(\ref{generalODEFixTime}) numerically with the same choice of $\bar{A}(z)$ and $\bar{B}(z)$.

We find that simulation and the one-dimensional approximation give excellent agreement for a wide range of parameters, as demonstrated in \fref{smallS1}. For systems in which the form of $a_{1}$ and $b_{1}$ result in a large selective advantage for one or the other of the alleles, the time to fixation can be observed to clearly lose the symmetric form observed in the neutral case. This can be seen in \fref{smallS1} for the results represented in green/squares and those in blue/circles.

%%%%%%%%%%%%%%%%%%%%%%%%%%%%%%%%%%%%%%%%%%%%%%%%%%%%%%%%%%%%%%%%%%%%%%%%%%%%%
%%%%%%%%%%%%%%%%%%%%%%%%%%%%%%%%%%%%%%%%%%%%%%%%%%%%%%%%%%%%%%%%%%%%%%%%%%%%%

\begin{figure}[th]
 \includegraphics[width=0.45\textwidth]{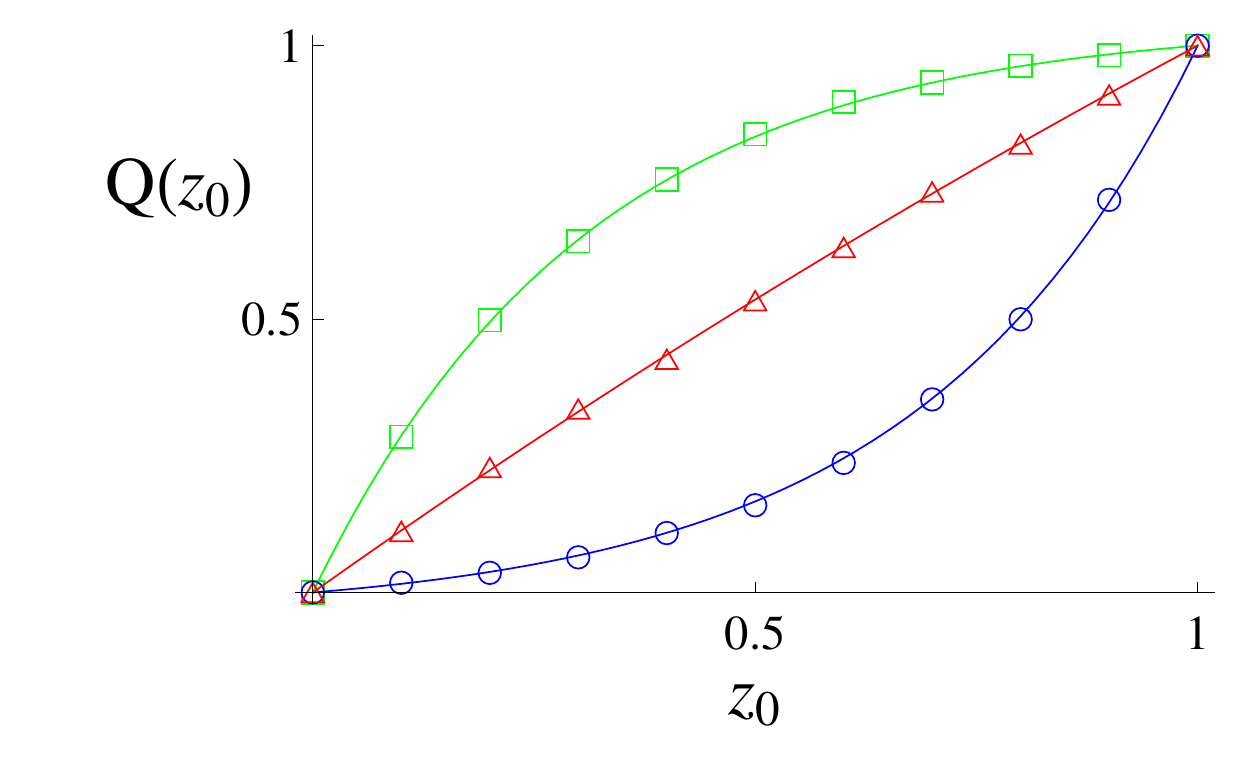}
 \includegraphics[width=0.45\textwidth]{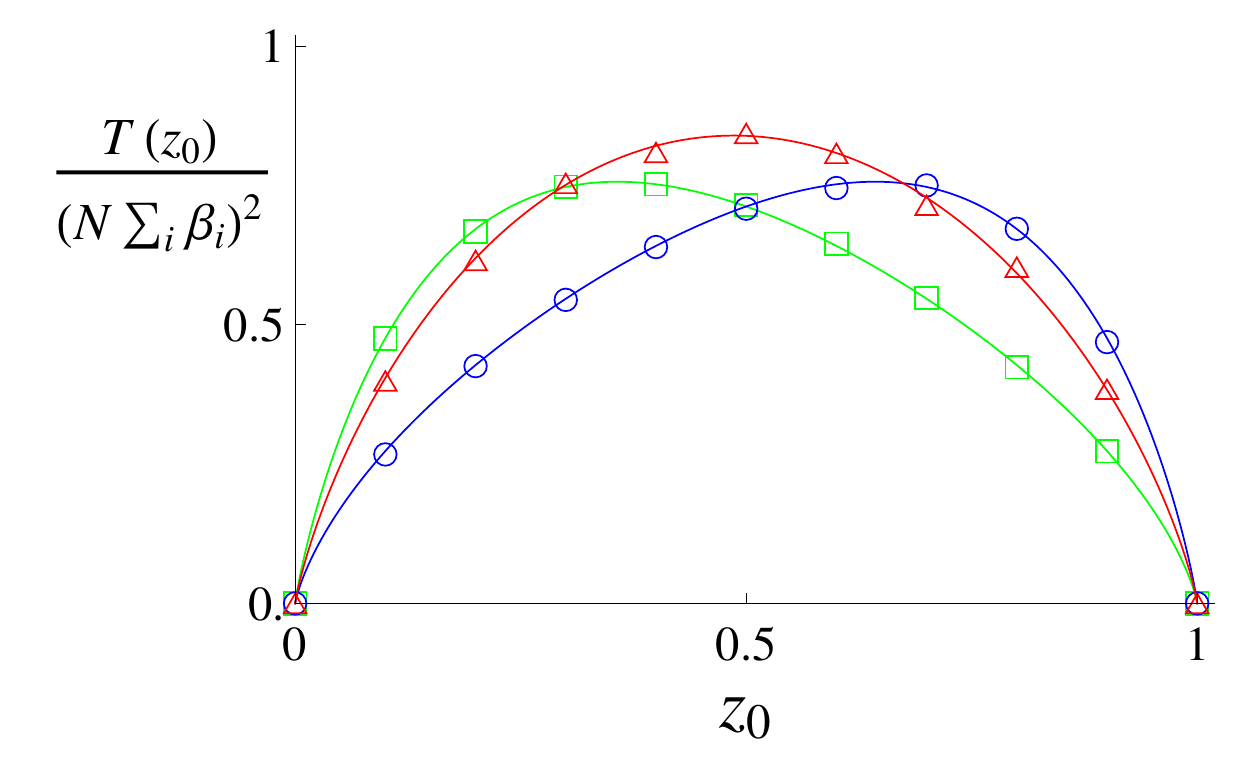}
\caption{(Color online) Plots for the probability of fixation (upper panel) and mean time to fixation (lower panel) as a function of the projected initial conditions for the first order in $s$ case. Continuous lines are obtained from \Eref{Q_first_order} and by solving \eref{generalODEFixTime} numerically with $\bar{A}(z)$ to order $s$ and $\bar{B}(z)$ given by \Eref{defineBbar}. Various symbols indicate the results obtained from simulations. For each different color/symbol different $\bm{\alpha}$ vectors are used; green squares, $\bm{\alpha}=(1,1,-1)$, red triangles, $\bm{\alpha}=(-1,-1,1)$, and blue circles, $\bm{\alpha}=(1,-2,-1)$. All other parameters are kept constant; $s=0.005$, $N=200$, $\bm{\beta}=(3,2,1)$ and the migration matrix $m$ is fixed, though not given here. Simulation results are the average of $5000$ runs.}
\label{smallS1}
\end{figure}

%%%%%%%%%%%%%%%%%%%%%%%%%%%%%%%%%%%%%%%%%%%%%%%%%%%%%%%%%%%%%%%%%%%%%%%%%%%%%
%%%%%%%%%%%%%%%%%%%%%%%%%%%%%%%%%%%%%%%%%%%%%%%%%%%%%%%%%%%%%%%%%%%%%%%%%%%%% 

We now perform a similar comparison in the case where $\bar{A}(z)$ truncated at second order in $s$. While analytical results for the probability of fixation and mean time to fixation can be obtained in this case, as they can be in the case when $\bar{A}(z)$ is linear in $s$, they will be discussed elsewhere~\cite{projectionBio}. Here we restrict ourselves to solving the differential equations \eref{generalODEFixProb} and \eref{generalODEFixTime} numerically in order to demonstrate the predictive power of the reduced model. We will also choose model parameters which are not related to any specific application for a similar reason; the method in specific contexts will also be described elsewhere~\cite{projectionBio}. In \fref{QsN2} results are plotted for two particular parameter sets, in which very good agreement between simulation and theory is observed. In particular we note the plot for the probability of fixation, $Q(z_{0})$, in represented by blue lines and circles; the functional form observed here is unobtainable from the order $s$ description, \Eref{Q_first_order}. 

%%%%%%%%%%%%%%%%%%%%%%%%%%%%%%%%%%%%%%%%%%%%%%%%%%%%%%%%%%%%%%%%%%%%%%%%%%%%%
%%%%%%%%%%%%%%%%%%%%%%%%%%%%%%%%%%%%%%%%%%%%%%%%%%%%%%%%%%%%%%%%%%%%%%%%%%%%%

\begin{figure}
\includegraphics[width=0.45\textwidth]{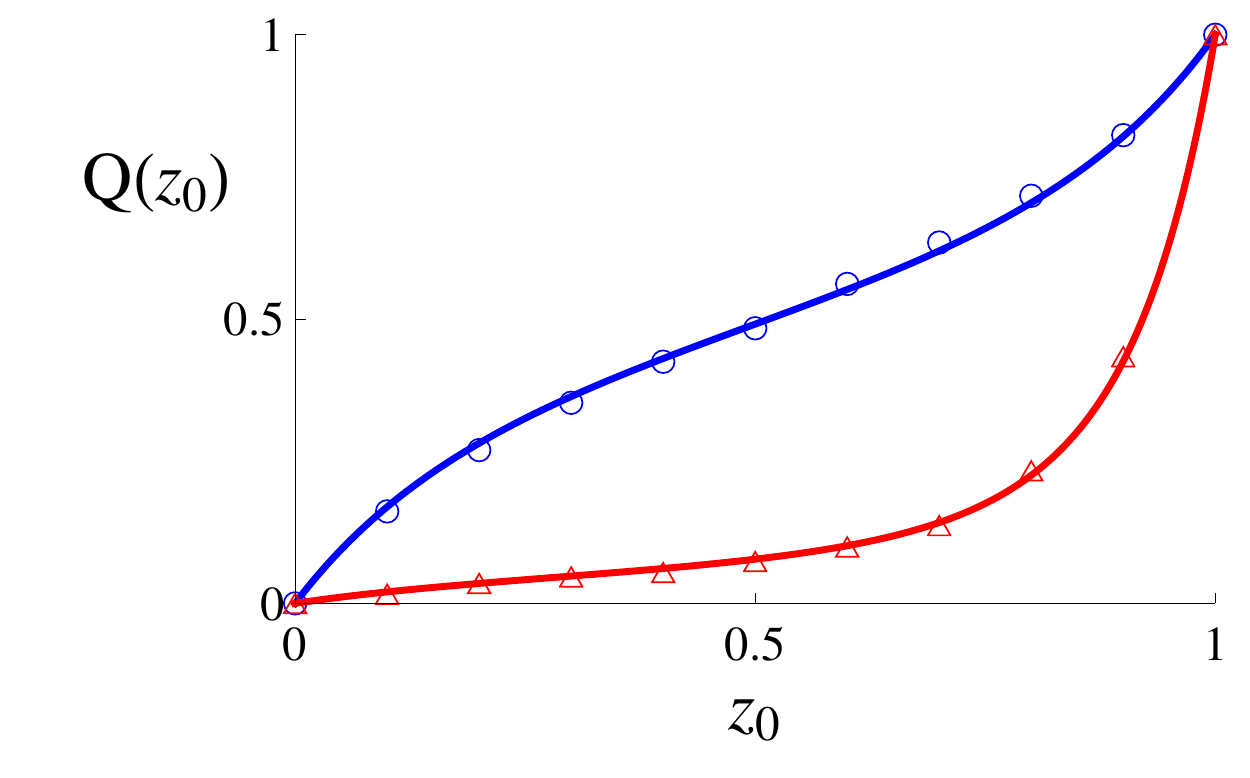}
\includegraphics[width=0.45\textwidth]{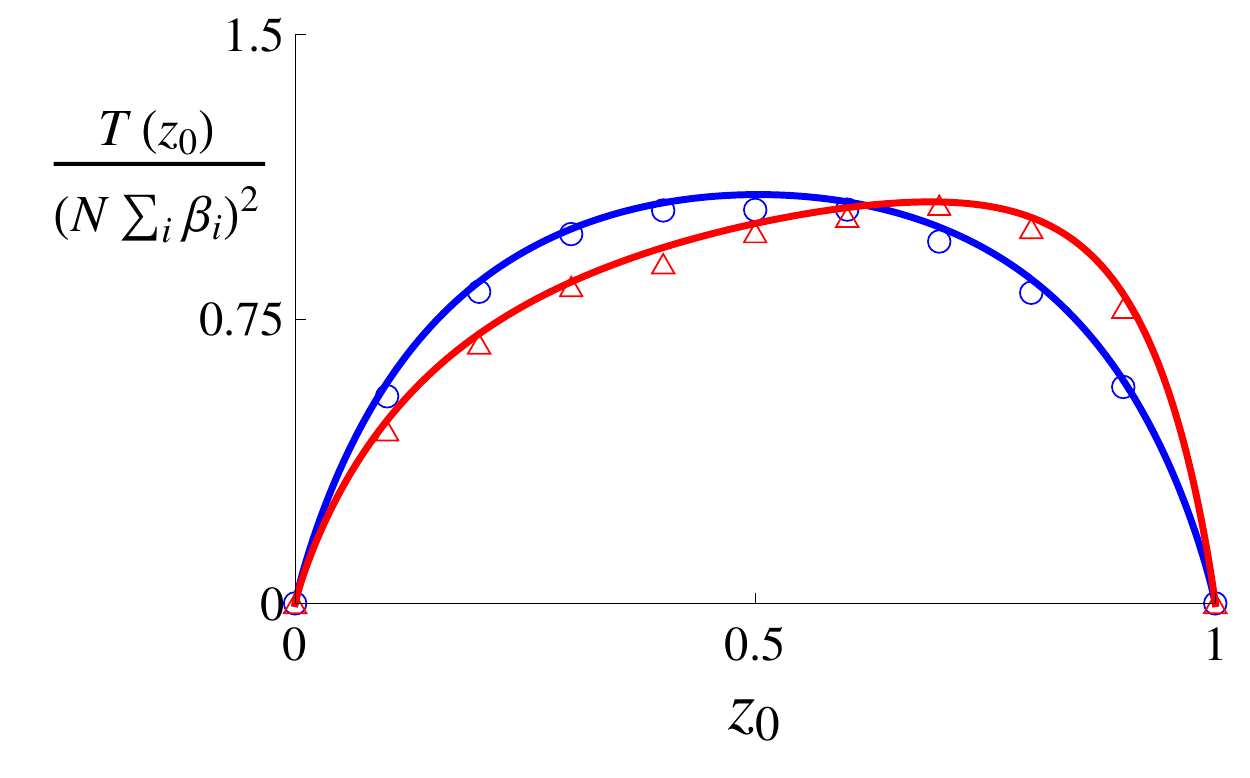}
\caption{(Color online) Plots for the probability of fixation, $Q(z_{0})$, and mean time to fixation, $T(z_{0})$, to second order in $s$. Continuous lines are obtained from the reduced model, solving \eref{generalODEFixProb} and \eref{generalODEFixTime} respectively. Parameters used for the results in blue circles are $\mathcal{D}=2$, $N=400$, $s=1/\sqrt{N}$, $\bm{\beta}=(1,1)$, $\bm{\alpha}=(1,-1)$. Parameters used for the results in red triangles are $\mathcal{D}=4$, $N=300$, $s=1/\sqrt{N}$, $\bm{\beta}=(1,1,1,2)$, $\bm{\alpha}=(1,1,0.3,-1)$. In both cases the migration matrices are non-symmetric. Simulation results are the mean of $10^{4}$ runs.}\label{QsN2}
\end{figure}

%%%%%%%%%%%%%%%%%%%%%%%%%%%%%%%%%%%%%%%%%%%%%%%%%%%%%%%%%%%%%%%%%%%%%%%%%%%%%
%%%%%%%%%%%%%%%%%%%%%%%%%%%%%%%%%%%%%%%%%%%%%%%%%%%%%%%%%%%%%%%%%%%%%%%%%%%%%

\section{Estimating the range of validity of the method}\label{secLimits}

Having discussed the approximation method and results, we now turn to considering the range of validity of the expressions for the reduced system, providing a heuristic argument along with a numerical analysis. 

As stated in \Sref{secNeutral} the quantities that govern the separation of timescales are the eigenvalues of $H$. While the first eigenvalue, $\lambda^{(1)}$, is always zero, we require that the remaining eigenvalues are sufficiently less than zero so that the collapse of the system onto the slow subspace or center manifold happens on a much faster timescale than that of fixation. What, then, is a sufficient separation of eigenvalues? To investigate this we consider a $\mathcal{D}=3$ system where each deme is of equal size, whose migration matrix is characterized by a single number $0<\theta<1$: 
\begin{eqnarray}
 m = \left( \begin{array}{ccc} \theta       & (1-\theta)/2 & (1-\theta)/2 \\
                        (1-\theta)/2 &    \theta    & (1-\theta)/2 \\
                         (1-\theta)/2& (1-\theta)/2 & \theta
     \end{array} \right) \, . \label{mBreakdown}
\end{eqnarray}
With these properties, the $H$ matrix for the system can be simply constructed. We find a degenerate system with only two eigenvalues, the first, zero, and the other two given by $\lambda^{(2)} = (\theta-1)/2$. We can then plot how the predictions of the reduced system, \Eref{generalSDE1D}, compare against the results of simulation for some fixed initial condition as $\lambda^{(2)}$ increases. 

The results in the neutral case are shown in \Fref{breakdown_s_0}. We recall that since the matrix is symmetric, $b_{1}$ is given by \Eref{defineb1_symmetric}. One can see that the reduced system agrees well with the probability of fixation across a remarkably large range of eigenvalues. While the prediction for the time to fixation fares slightly less well, results from simulation still agree over a very large range of parameters, only beginning to diverge at approximately $\lambda_{2}=-0.05$, at which point one begins to see an rapid increase in fixation time of the simulations. This is also the point at which the magnitude of the noise, moderated by $1/\sqrt{N}$, is of the order of the deterministic term (see \Eref{generalSDE}).

%%%%%%%%%%%%%%%%%%%%%%%%%%%%%%%%%%%%%%%%%%%%%%%%%%%%%%%%%%%%%%%%%%%%%%%%%%%%%
%%%%%%%%%%%%%%%%%%%%%%%%%%%%%%%%%%%%%%%%%%%%%%%%%%%%%%%%%%%%%%%%%%%%%%%%%%%%%

\begin{figure}
\includegraphics[width=0.45\textwidth]{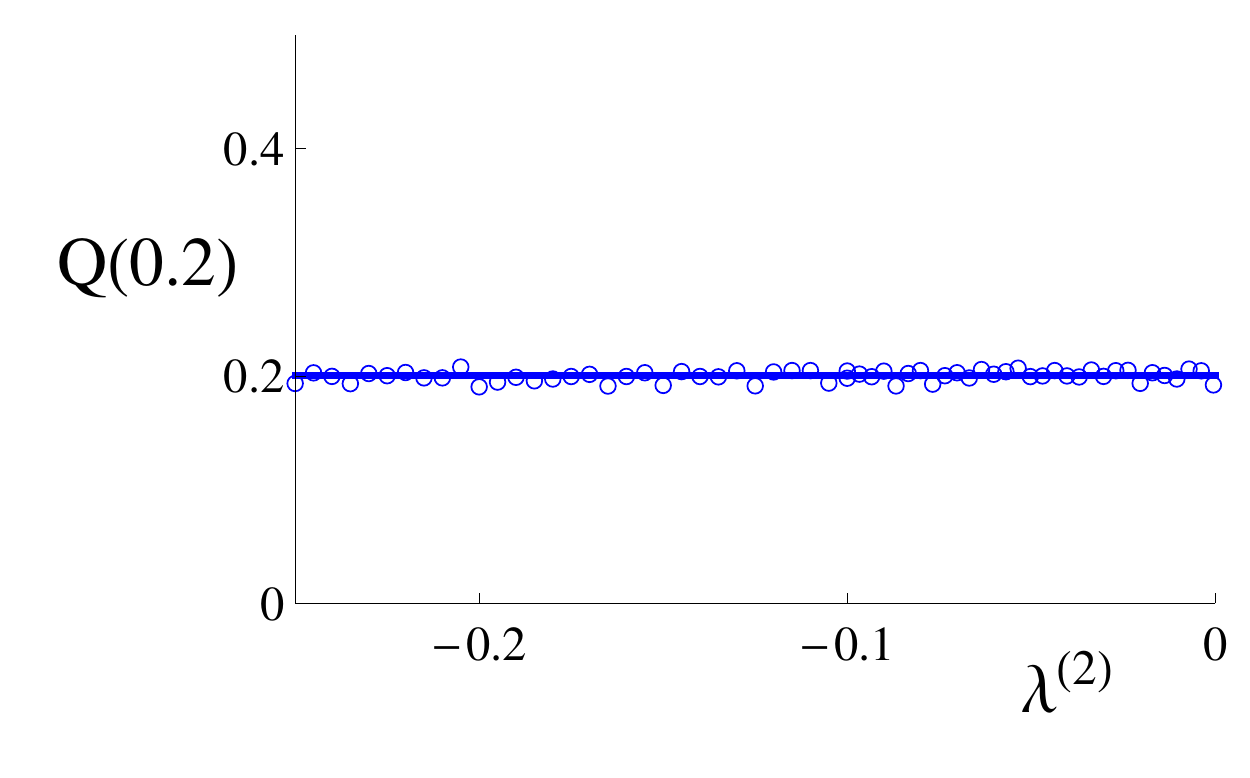}
\includegraphics[width=0.45\textwidth]{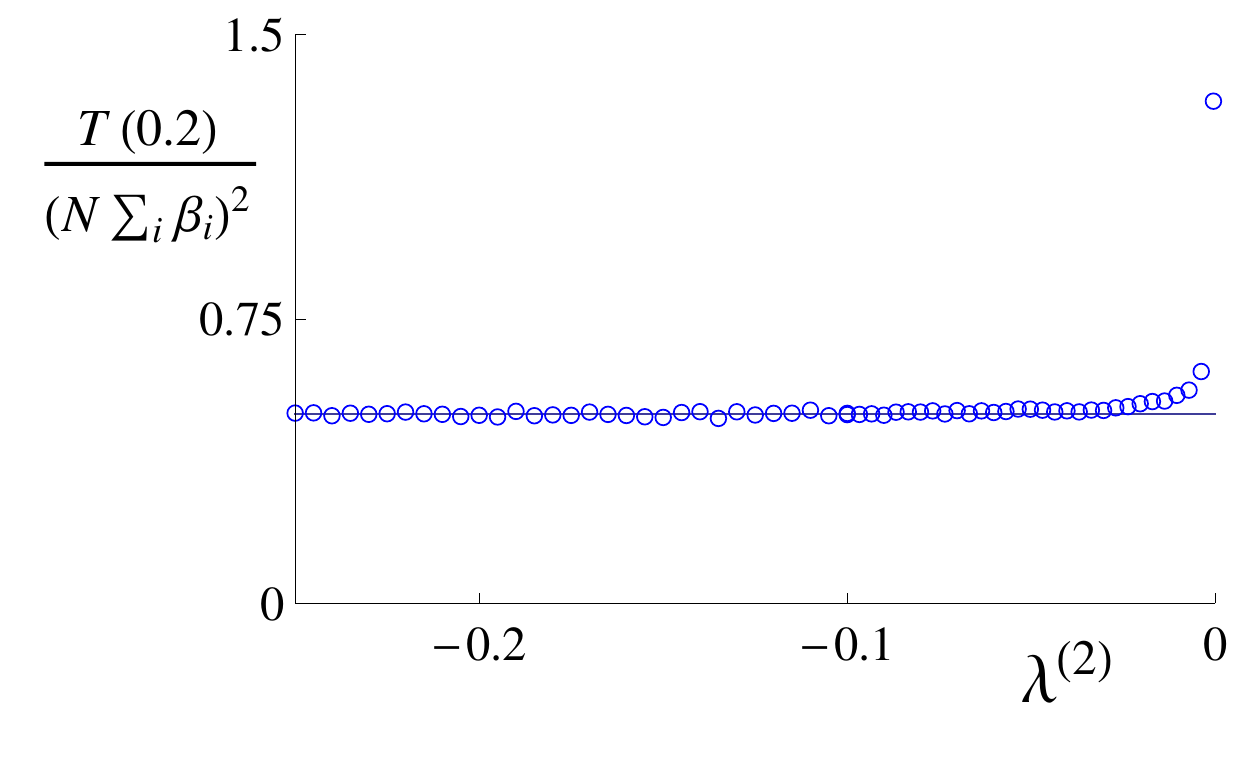}
\caption{(Color online) Plots of the probability of fixation and normalized time to fixation as a function of increasing eigenvalue (equivalent to decreasing the strength of collapse onto the center manifold). The system is prescribed by the migration matrix given in \Eref{mBreakdown}, with $\mathcal{D}=3$, $\bm{\beta}=(1,1,1)$, $s=0$, and $N=200$, and the initial condition for both plots is $z_{0}=0.2$. Mean values from $10^{4}$ stochastic simulations are plotted as circles, whereas continuous lines represent theoretical predictions. The final point on both plots is $\lambda^{(2)}=-5 \times 10 ^{-4}$.}\label{breakdown_s_0}
\end{figure}

%%%%%%%%%%%%%%%%%%%%%%%%%%%%%%%%%%%%%%%%%%%%%%%%%%%%%%%%%%%%%%%%%%%%%%%%%%%%%
%%%%%%%%%%%%%%%%%%%%%%%%%%%%%%%%%%%%%%%%%%%%%%%%%%%%%%%%%%%%%%%%%%%%%%%%%%%%%

In the case where selection is present, $s \neq 0$, we can conduct a study of the same system for a fixed set of selection parameters. We recall here, that the reduction techniques relies on the term $\sum_{j=1}^{\mathcal{D}}H_{ij}x_{j}$ in \Eref{A_second_order} inducing a near-deterministic, linear collapse and restriction of the system to the slow subspace. In this situation, where $s\neq0$, this assumption is not only broken by the noise but also the order $s$ nonlinear terms in \Eref{A_second_order}. We might therefore expect the reduced system to perform less well with decreasing $\lambda^{(2)}$ than the neutral case. While we find this is the case (see \Fref{breakdown_s_N}), the approximation still works very well up to $\lambda^{(2)}\approx -0.05$. At this point our reduced system under-predicts the probability of fixation and over-predicts the time to fixation, growing rapidly, much faster than the results from simulation would suggest.

%%%%%%%%%%%%%%%%%%%%%%%%%%%%%%%%%%%%%%%%%%%%%%%%%%%%%%%%%%%%%%%%%%%%%%%%%%%%%
%%%%%%%%%%%%%%%%%%%%%%%%%%%%%%%%%%%%%%%%%%%%%%%%%%%%%%%%%%%%%%%%%%%%%%%%%%%%%

\begin{figure}
\includegraphics[width=0.45\textwidth]{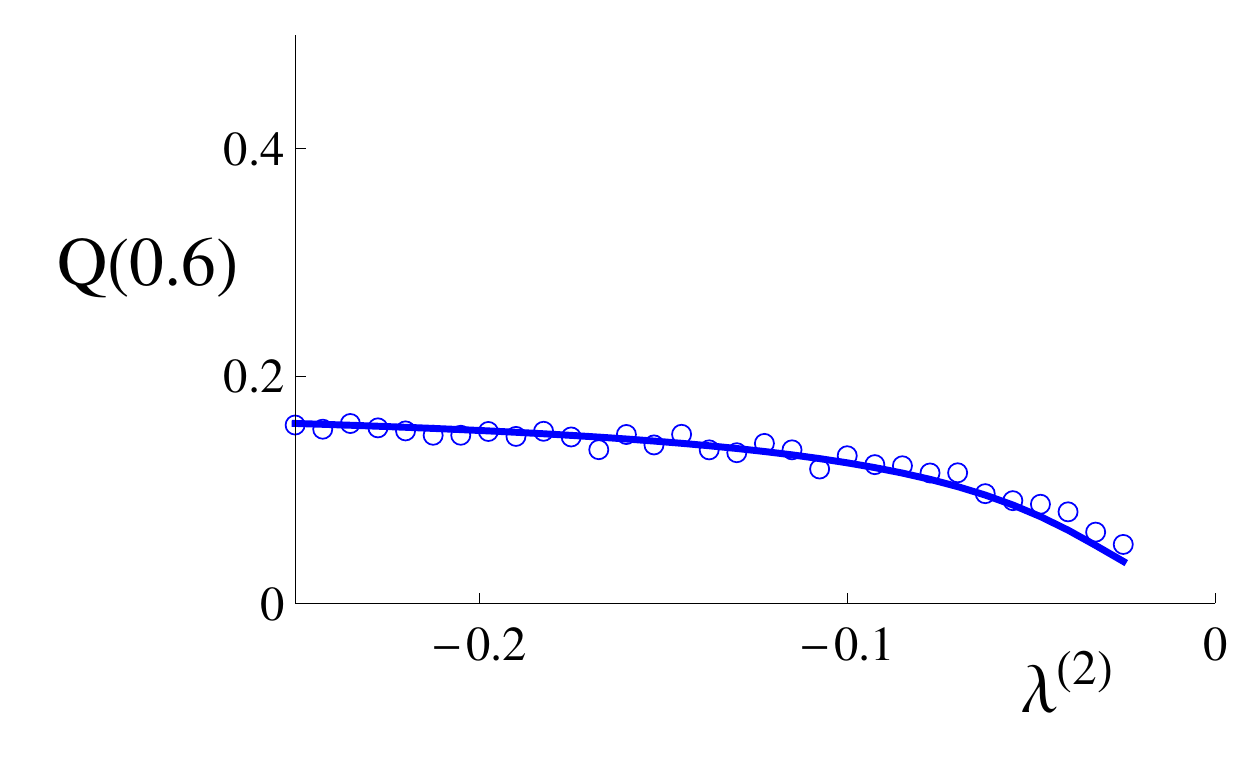}
\includegraphics[width=0.45\textwidth]{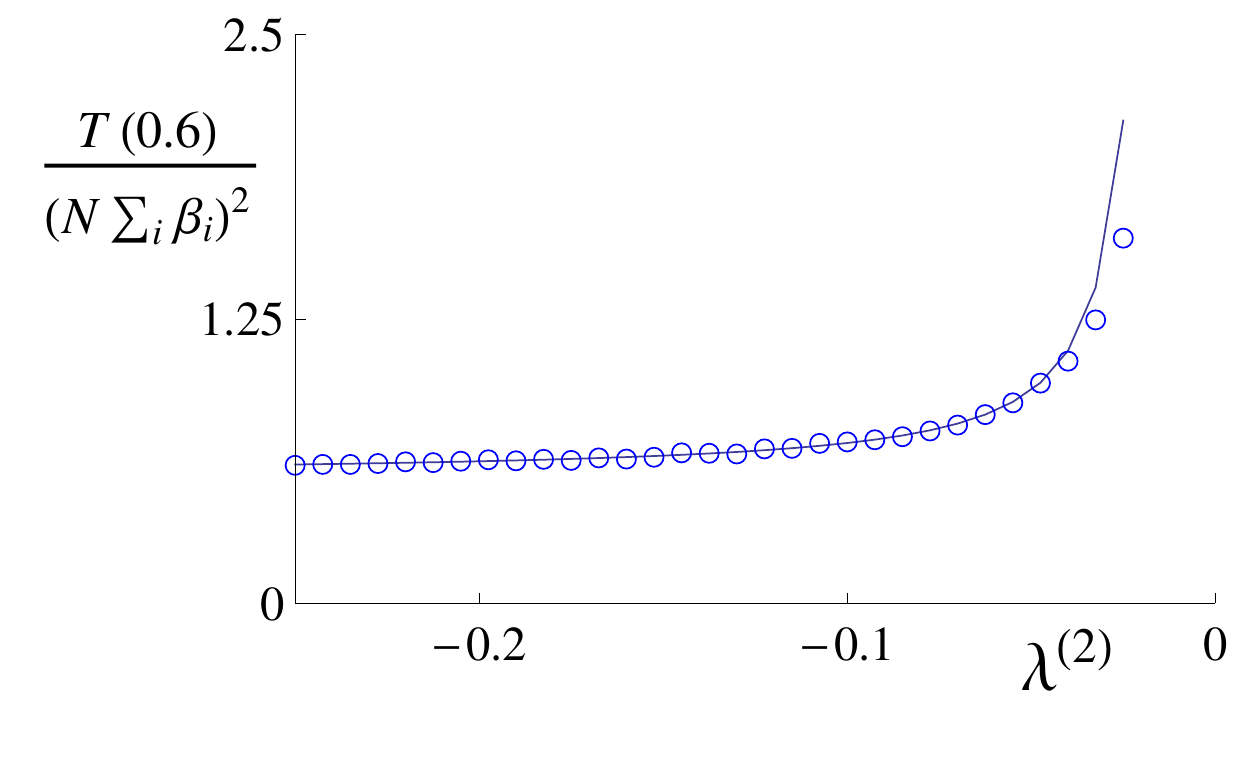}
\caption{(Color online) Plots of the probability of fixation and normalized time to fixation as a function of increasing eigenvalue in a system with $s = 0.035 $. The system is prescribed by the migration matrix given in \Eref{mBreakdown}, with $\mathcal{D}=3$, $\bm{\beta}=(1,1,1)$, $N=200$ and $\bm{\alpha}=(1,-2,0.5)$, and the initial condition for both plots is $z_{0}=0.6$. Mean values from $6 \times 10^{3}$ stochastic simulations are plotted as circles, whereas continuous lines represent theoretical predictions. The final point on both plots is $\lambda^{(2)}= -0.025 $.}\label{breakdown_s_N}
\end{figure}

%%%%%%%%%%%%%%%%%%%%%%%%%%%%%%%%%%%%%%%%%%%%%%%%%%%%%%%%%%%%%%%%%%%%%%%%%%%%%
%%%%%%%%%%%%%%%%%%%%%%%%%%%%%%%%%%%%%%%%%%%%%%%%%%%%%%%%%%%%%%%%%%%%%%%%%%%%%

\section{Conclusions}\label{secConclusion}

The methodology of statistical mechanics is frequently applicable in areas outside of the physical sciences in cases where a large number of entities of a similar type interact with each other. However, the nature of the interaction will usually be more varied than in physics or chemistry. One kind of system structure that is often encountered is that of a set of `islands', each containing a large number of constituents, which interact with each other with a strength given by the magnitude of the links in a fixed network. The difficulty in studying the statistical dynamics of such systems is the complexity of the governing equations. Even after making a diffusion approximation, they will consist of a set of coupled non-linear SDEs with multiplicative noise.

In this paper we have presented a method of reducing this set of SDEs down to a single SDE, which can then be analyzed. The method was applied in the specific case of population genetics where the islands are demes, the constituents are individuals with a gene which can be one of two types, and with the interaction being the migration between demes. The approximation was based on the elimination of fast degrees of freedom, which resulted in a stochastic dynamics that was confined to a one-dimensional slow subspace. We expect this method to be applicable to many other areas, such as those referred to the \Sref{secIntroduction}, but chose to illustrate the idea in the case of population genetics because of the importance of selection in that context.

The method we have developed is in excellent agreement with simulations, is simple to understand, and results in a what is effectively an algorithm which may be applied to any network. The constants $a_1,a_2$ and $a_3$ which appear in the drift term in the final SDE, \Eref{ABarSNHalf}, are given by precise formulas which can be calculated given any network structure. These are the constants that appear when the form of the SDE is calculated to order $s^2$, where $s$ is the selection strength. It is possible to go to higher order in $s$, but $s$ is usually so small that this would almost certainly never be required; studies which work to order $s^2$ are already extremely rare, most authors being content to work only to order $s$. 

The method does not assume that the number of individuals is the same in each deme, although we expect the results will be less reliable if some demes are an order of magnitude bigger than others. If this is the case, the elimination of fast modes may still go through, but different methods of approximation would be more applicable. Similarly, as discussed in \Sref{secLimits}, if the non-zero eigenvalues of the matrix $H$ are too close to zero, then the approximation may break down. This means that, a priori, one cannot tell if the reduction technique is applicable for general $m$, $\bm{f}$ and $s$. Instead one must first construct the matrix $H$ to determine its eigenvalues. One can then check that the parameters lie broadly in the range identified in \Sref{secLimits}; the non-zero eigenvalue of $H$ which is closest to zero, $\lambda^{(2)}$,  should have a magnitude greater than both $\mathcal{O}(N^{-1/2})$ and $s$. The mathematical origins of these limits are interesting points to investigate in future work. 

Here we have confined ourselves to describing the methodology of the reduction from many coupled SDEs to a single SDE, and to showing that it is an excellent approximation for a wide range of parameters. We will explore the consequences for a number of problems in population genetics elsewhere \cite{projectionBio}, but we believe that the method is quite general, and hope and expect that it will be used in a large range of applications in the future.

\begin{acknowledgments}
We thank Tim Rogers for useful discussions. G.W.A.C. thanks the Faculty of Engineering and Physical Sciences, University of Manchester for funding through a Dean's Scholarship.
\end{acknowledgments}

\begin{appendix}

\section{Calculation of the dynamics on the slow subspace}
\label{appSlowManifold}
In this appendix some of the more technical aspects of finding the slow subspace and calculating the dynamics of the reduced system will be set out. So far we have only specified the natural variable which we use in the reduced system, that is $z = \sum_{i=1}^{\mathcal{D}} u^{(1)}_{i}x_{i}$. 

More generally, we can define a linear transformation to the coordinate $z$ and $\mathcal{D}-1$ coordinates $\bm{y}$ such that 
\begin{eqnarray}
\left( \begin{array}{c} z \\ \bm{y} \end{array} \right) = T^{-1} \bm{x} \, , \quad \bm{x} = T \left( \begin{array}{c} z \\ \bm{y} \end{array} \right) \, .
\end{eqnarray}
A convenient choice for $x_{i}$ is 
\begin{eqnarray}\label{generalX}
 x_{i} = z + \sum_{a=1}^{\mathcal{D}-1} W_{ia}y_{a} \,.
\end{eqnarray}
Since, from Eq.~(\ref{define_center}), $x_i = z$ on the center manifold in the neutral case, we ask that the $y_a$ are of order $s$ on the slow subspace in the case with selection. This will simplify our calculation because, as we will see, this means that we will only have to calculate the $y_a$ as functions of $z$ to leading order in $s$.

In terms of the transition matrix $T$, the choices made so far mean that
\begin{eqnarray}
 T^{-1} = \left( \begin{array}{c} [\bm{u}^{(1)}]^{T} \\ R \end{array} \right) \,, \quad T = \left( \bm{1} \quad W \right) \, ,
\end{eqnarray}
where $R$ is a $\mathcal{D}-1$ by $\mathcal{D}$ matrix and $W$ is a $\mathcal{D}$ by $\mathcal{D}-1$ matrix. The form of the matrices $R$ and $W$ is restricted through the conditions $T T^{-1} = T^{-1}T = I$, the identity matrix. The condition relevant if we are trying to express $\bm{x}$ in terms of $z$ and $\bm{y}$, is 
\begin{eqnarray}\label{inverseConditions}
\sum_{i=1}^{\mathcal{D}} u^{(1)}_{i}W_{ia} = 0 \,, \ \ a=1,\ldots,\mathcal{D}-1.
\end{eqnarray}
We will need to check that any choice we make for $W_{ia}$ satisfies this condition.

We now substitute the transformation (\ref{generalX}) into Eq.~(\ref{A_second_order}) for the drift vector in terms of $\bm{x}$:
\begin{eqnarray}\label{generalFitnessDriftZY}
& & A_{i}(z, \bm{y}) = \sum_{j=1}^{\mathcal{D}}H_{ij}
\left(  z  + \sum_{a=1}^{\mathcal{D}-1} W_{ja}y_{a} \right)\nonumber \\
		    &+& s z (1 - z) \sum_{j=1}^{\mathcal{D}}
\frac{G_{ij}\alpha_j}{\beta_{i}} \nonumber \\
		      &+& s(1-2z) \sum_{j=1}^{\mathcal{D}} 
\frac{G_{ij}\alpha_j}{\beta_{i}} 
\sum_{a=1}^{\mathcal{D}-1} W_{ja}y_{a} \nonumber \\
			&-& s^{2}z^{2}(1-z) 
\sum_{j=1}^{\mathcal{D}} \frac{G_{ij}\alpha^2_j}{\beta_{i}}
+ \mathcal{O}(s^{2}y, s^{3}) \, .
\end{eqnarray}
Using (i) $\sum^{\mathcal{D}}_{j=1} H_{ij}z = z \sum^{\mathcal{D}}_{j=1} H_{ij} = 0$, from Eq,~(\ref{defineH}), and (ii) the slow subspace condition $\sum^{\mathcal{D}}_{i=1}u^{(a+1)}_{i}A_{i} = 0$, $a=1,\ldots,\mathcal{D}-1$ (see Eq.~(\ref{centerManifold})), we find
\begin{eqnarray}\label{slowManifoldCondition}
0 &=& \sum_{i,j=1}^{\mathcal{D}}\sum^{\mathcal{D}-1}_{a=1} u^{(a+1)}_{i} H_{ij} W_{ja} y_a \nonumber \\
&+& s z (1 - z) \sum_{i,j=1}^{\mathcal{D}}\frac{u^{(a+1)}_{i}G_{ij}\alpha_j}{\beta_{i}}\,,
\end{eqnarray}
since the slow subspace condition must be satisfied order by order in $s$ and $\bm{y}$ is assumed to be of order $s$. Choosing $W_{ja}$ to be the right-eigenvectors $v^{(a+1)}_{j}$, $a=1,\ldots,\mathcal{D}-1$, which is consistent with the conditions (\ref{inverseConditions}), we see that the first term on the right-hand side of Eq.~(\ref{slowManifoldCondition}) is simply $\lambda^{(a+1)}y_a$. Therefore
\begin{eqnarray}
y_{a}(z) = - \frac{s z (1 - z) }{\lambda^{(a+1)}} \sum_{i,j=1}^{\mathcal{D}}\frac{u^{(a+1)}_{i}G_{ij}\alpha_j}{\beta_{i}} + \mathcal{O}(s^{2}) \,.
\label{y_intermsof_z}
\end{eqnarray}

Substituting Eq.~(\ref{y_intermsof_z}) into Eq.~(\ref{generalFitnessDriftZY}), the drift vector evaluated on the slow subspace is found to be
\begin{eqnarray}\label{generalFitnessDriftSlowManifold}
 A_{i}(z) = &-& s q^{(0)}_{i} z (1 - z) + s q^{(1)}_{i} z(1 - z)  - s^{2} q^{(2)}_{i} z^{2}(1 - z) \nonumber \\ &-& s^{2} q^{(3)}_{i} z (1 - z) (1 - 2 z) + \mathcal{O}(s^{3}) \,,
\end{eqnarray}
where the vectors $\bm{q}^{0}$, $\bm{q}^{1}$, $\bm{q}^{2}$ and $\bm{q}^{3}$ are the parameter combinations
\begin{eqnarray}
q^{(0)}_{i} &=& \sum^{\mathcal{D}-1}_{a=1} \sum^{\mathcal{D}}_{j,k,l=1} \frac{H_{ij}v^{(a+1)}_{j}u^{(a+1)}_{k}G_{kl}\alpha_{l}}{\beta_{k}\lambda^{(a+1)}}, \nonumber \\
q^{(1)}_{i} &=& \sum_{j=1}^{\mathcal{D}} \frac{G_{ij}\alpha_j}{\beta_{i}}\,, \ \ \ \
q^{(2)}_{i} = \sum_{j=1}^{\mathcal{D}} \frac{G_{ij}\alpha^2_j}{\beta_{i}}\,, 
\nonumber \\
q^{(3)}_{i} &=& \sum_{a=1}^{\mathcal{D}-1}\sum_{j,k,l=1}^{\mathcal{D}}
\frac{G_{ij}\alpha_j}{\beta_{i}} 
\frac{v^{(a+1)}_{j} u^{(a+1)}_{k}}{\lambda^{(a+1)}} 
\frac{G_{kl}\alpha_l}{\beta_{k}}.
\label{qs}
\end{eqnarray}

The elements of the diffusion matrix meanwhile, when evaluated on the slow subspace, have the form
\begin{eqnarray}\label{generalFitnessDiffusionSlowManifold}
 B_{ii}(z) = 2 z (1 - z) \sum_{j=1}^{\mathcal{D}}\frac{ G_{ij} } {\beta_{i}^{2}} + \mathcal{O}(s) \, .
\end{eqnarray}

Since the matrix $H$ is not in general symmetric, then the eigenvalues will not in general be real. However since the entries of $H$ are real, the eigenvalues will occur in complex conjugate pairs, and the eigenvectors associated with an eigenvalue $\lambda^{*}$ will be the complex conjugates of those associated with $\lambda$. Since the expressions for $q^{(0)}_i$ and $q^{(3)}_i$ in Eq.~(\ref{qs}) take the form of sums over $a$, for each term which is not real there will be another term added to it which is its complex conjugate. Thus $q^{(0)}_i$ and $q^{(3)}_i$ are guaranteed to be real. Therefore, the procedure goes through whether the eigenvalues are real or not. Of course, if there are complex conjugate pairs, the corresponding $y_a$ cannot be interpreted as coordinates. However this interpretation is not crucial to the method, and if one wishes, it is always possible to define real coordinates by working with the real and imaginary parts of the eigenvalues and eigenvectors.

\end{appendix}

%\bibliographystyle{plain}
%\bibliography{popPhys5}

%merlin.mbs 2010-03-15 4.21a (PWD, AO, DPC)

\end{document}